\documentclass[usenatbib]{mnras}
\usepackage[T1]{fontenc}
\usepackage{ae,aecompl}
\usepackage{setspace}
\usepackage{parskip}
\usepackage{amsmath}
\usepackage{amssymb}
\usepackage{mathtools}
\usepackage{graphicx}
\usepackage{subcaption}
\usepackage{array}
\usepackage{enumerate}
\usepackage{verbatim}
\usepackage[english]{babel}
\usepackage{siunitx}
\usepackage{multirow}

\DeclareSIUnit\jansky{Jy}
\DeclareSIUnit\erg{erg}

\makeatletter
\newcommand{\bianca}{\renewcommand\NAT@open{[}\renewcommand\NAT@close{]}}
\makeatother

\newcommand*\sqcitet[1]{{\bianca\citet{#1}}}
\newcommand*\sqcitep[1]{{\bianca\citep{#1}}}

\title[Narrow, Intrinsic C\,{\normalsize \textit{IV}} Absorption in Quasars]{Narrow, Intrinsic C\,{\Large \textbf{IV}} Absorption in Quasars as it Relates to Outflows, Orientation, and Radio Properties}

\author[R. B. Stone et al.]{
Robert B. Stone,$^{1}$\thanks{E-mail: robert.b.stone@drexel.edu}
Gordon. T. Richards,$^{1}$
\\
$^{1}$Department of Physics, Drexel University, 32 S.\ 32nd Street, Philadelphia, PA 19104 USA
}

\date{Accepted XXX. Received YYY; in original form ZZZ}

\pubyear{2019}

\begin{document}
\label{firstpage}
\pagerange{\pageref{firstpage}--\pageref{lastpage}}
\maketitle

\begin{abstract}
This work provides evidence that a large fraction of \ion{C}{IV} narrow absorption lines (NALs) seen along the line of sight to distant quasars are due to accretion disk winds, while also seeking to understand the relationship between NALs and certain quasar-intrinsic properties.  We extend the results from past work in the literature using $\sim105,000$ NALs from a sample of $\sim58,000$ SDSS quasars.  The primary results of this work are summarized as follows:  (1) the velocity distribution (${\rm d}N/{\rm d}\beta$) of NALs is not a function of radio loudness (or even detection) once marginalized by optical/UV luminosity;  (2) there are significant differences in the number and distribution of NALs as a function of both radio spectral index and optical/UV luminosity, and these two findings are not entirely interdependent; (3) improvements in quasar systemic redshift measurements and differences in the NAL distribution as a combined function of optical luminosity and radio spectral index together provide evidence that a significant portion of NALs are due to outflows; (4) the results are consistent with standard models of accretion disk winds governed by the $L_{\rm UV}$-$\alpha_{ox}$ relationship and line-of-sight orientation indicated by radio spectral index, and (5) possibly support a magnetically arrested disk model as an explanation for the semi-stochastic nature of strong radio emission in a fraction of quasars.
\end{abstract}


%%%%%%%%%%%%%%%%%%%%%%%%%%%%%%%%%%%%%%%%%%%%%%%%%%

%%%%%%%%%%%%%%%%% BODY OF PAPER %%%%%%%%%%%%%%%%%%

\section{Introduction} \label{sec:intro}

Quasar absorption lines systems (QSOALSs) provide great insight both to the inner workings of quasars and to galaxies, gas, and dust along their line of sight.  Strong QSOALSs have two main sources:  1) Matter associated with the quasar and/or its host galaxy (e.g., molecular clouds, nearby dwarf galaxies, winds from the accretion disk, or galactic-scale outflows/inflows) and 2) intervening material along the line of sight (e.g., galaxies and filaments of hydrogen gas).

QSOALSs are further split into two classes based on their line widths:  broad absorption lines (BALs) and narrow absorption lines (NALs)---although there is no clear demarcation between the two and there exist intermediate systems that are referred to as ``mini-BALs" \citep{Turnshek1988,Barlow1997}.  BALs, formally defined by the \citet{Weymann1991} criteria (or balnicity index; BI), are largely considered to be associated with outflowing matter from the central engine \citep[e.g.,][]{Weymann1991,Richards2006a,Ganguly2007}.  NALs, on the other hand, with equivalent widths of just a few hundred \si{\km\per\s}, can either be associated with the host or be intervening systems \citep{Weymann1979,Foltz1986,Richards2001b,Wild2008}.  NALs are seen in about 60\% of quasar spectra \citep{Vestergaard2003,Ganguly2008}, while BALs are seen in only roughly 20\% \citep[e.g.,][]{Hewett2003,Knigge2008}.  NALs that are within about \SI{3000}{\km\per\s} of the systemic redshift of the QSO are termed ``associated absorption lines" (AALs).

Based on the redshift of a QSOALS alone, it is not possible to know whether a NAL system represents intrinsic or intervening material and, at first, it would seem that NALs are distinct from broad absorption lines.  However, a number of investigations have clearly shown that some NALs are intrinsic \citep{Weymann1979,Foltz1986,Richards2001b,Vestergaard2003,Misawa2007,Wild2008,Bowler2014}, particularly those that have redshifts that are close to the systemic redshift of the quasar: $z_{\rm abs} \sim z_{\rm qso}$.  Traditionally thought to be caused by material trapped in the host QSO's potential well \citep[e.g.,][]{Weymann1979,Foltz1986}, more recently it has been suggested that (at least some) AALs are the result of an outflow from the accretion disk, perhaps in a direction at some angle to our line-of-sight \citep{Richards2001b,Vestergaard2003}.  It is even less clear in the case where there is a large difference between the quasar redshift and the NAL absorber redshift; such systems are typically intervening, but they can also be a fast-moving outflow \citep{Hamann2011,Bowler2014}.  To better understand outflows from quasars as a whole, we need to understand the extent to which some NAL systems are outflows and what their properties are.

If all the systems are intervening, then the metric that we are interested in is the number density per unit redshift ${\rm d}N/{\rm d}z$; however, if the systems are all intrinsic (whether outflows or inflows), then what we want to know is ${\rm d}N/{\rm d}\beta$, the number density per unit velocity, where $\beta = v_{\rm abs}/c$.  If we make the temporary assumption that all QSOALSs are intrinsic, then splitting up the QSOALSs in broad velocity bins can help unravel (at least statistically) the origin of the lines.  Specifically, strong QSOALSs can be described in three classes based on their velocity distribution:
\begin{enumerate}[(i)]
	\item Systems within a few thousand \si{\km\per\s} ($\lesssim \SI{3000}{\km\per\s}$), called associated absorption lines (AALs) \citep{Weymann1979,Wild2008}.
	\item Systems with velocity separations between $\sim 3000$ and \SI{12000}{\km\per\s}, dominated by systems intrinsic to and outflowing from the QSO.
	\item Systems with velocity differences $\gtrsim \SI{12000}{\km\per\s}$ that are predominantly spatially disconnected from the quasar (but see, e.g., \citealt{Hamann2011}).
\end{enumerate}
These three components can be seen in Figure~1 of \citet{Weymann1979}, and are displayed in Section~\ref{sec:beta_dist_z} of this paper in the context of our data sample, where we demonstrate the effect that improved QSO systemic redshift estimations have on the velocity distribution.

While in some cases it is possible to determine if a NAL is intrinsic or not (partial covering: e.g., \citealt{Hamann2011}; resolved absorption: for example mini-BALs; an unlikely statistical overdensity in a single object: e.g., \citealt{Richards2002}; or linelocking e.g., \citealt{Bowler2014}), it is also possible to draw broad statistical conclusions about the population of NALs as a whole.  Thus, our goal herein is not to specifically determine which NALs are due to outflows and which are due to intervening galaxies, but rather to use intrinsic properties of quasars (which should be independent of intervening NALs) to learn something about the frequency and properties of intrinsic NALs and to use that information to achieve a broader understanding of quasar outflows that transcends the traditional BAL criteria. 
 
In particular, past work has found strong differences in the distribution of NALs as a function of radio spectral index \citep{Foltz1986,Anderson1987,Foltz1988,Richards2001b} and optical luminosity \citep{Moller1987} and possibly between NALs hosted by radio-loud and radio-quiet quasars \citep{Foltz1986,Anderson1987,Foltz1988,Richards2001b}, however see \citet{Vestergaard2003} for a counter-example.  In this work we will first investigate the robustness of these claims using the largest yet sample of quasars and QSOALS to be investigated in this way.  We then look for additional properties that reveal differences in the NAL distribution, which would further shed light on the intrinsic/outflow population.

``Feedback" from the AGN to its host galaxy (and surrounding environment) can be recognized via outflowing absorption features (including the NALs discussed herein) and many models exist to explain such features.  They range from momentum- and energy-driven winds \citep[e.g.,][]{King2010,Faucher-Giguere2012b,Zubovas2014}, to radiation line driving (\citealt{Murray1995}, hereafter MCGV95; \citealt{Proga2000}), to radiation pressure on dust \citep{Thompson2005,Keating2012}.   To explain some (broad) absorption systems at large distances \citet{Faucher-Giguere2012a} rely on `in situ' radiative shocks in contrast to MCGV95 winds that may be more common for absorbers at smaller distances; see also \citet{Barthel2017}.  Broad reviews of the subject of AGN outflows/feedback can be found in \citet{Silk1998,Fabian1999,Begelman2004,King2015}.  

\citet[and references therein]{Proga2007} detail three sources of AGN winds: (1) thermal driving, (2) radiation pressure driving, and (3) magnetic driving.   Thermal driving happens when Compton heating from inner-disk X-rays create high enough tempteratures in the upper disk atmosphere to produce winds with greater velocity than the escape velocity \citep{Begelman1983,Proga2002}.  This process can be more common at large radii where the escape velocity is small, resulting in observed low-velocity X-ray absorption features \citep[e.g.,][]{Chelouche2005}.  Line-driving radiation pressure occurs when local disk radiation ejects gas (of particular ionized states) out of the accretion disk at radii where the disk mostly radiates in the UV (MCGV95; \citealt{Proga2000}).  This wind is continuous with velocities that can explain the observed blueshifted absorption lines in AGN spectra.  In this process, it is crucial that the gas not be too highly ionized, otherwise the fraction of ions that the radiation could exert pressure on would be too low to cause a continuous wind.  Radiation-driven AGN winds can be compared to winds from hot stars \citep{Lamers1999}, as illustrated by the P-Cygni-like troughs seen in BALs.  However, because of the threat of overionization of the gas by the hot X-ray flux, many radiation driving models include some form of self-shielding (MCGV95; \citealt{Proga2000,Leighly2004,Luo2015}).  Radiation driving appears to particularly dominate in the most luminous AGNs where the UV to X-ray flux ratio is largest \citep{Proga2000,Steffen2006}.  Finally, magnetically driven winds occur in two classes: (1) magnetocentrifugal winds and (2) magnetic pressure-driven winds.  The centrifugal force is shown to be able to drive a wind if the poloidal component of the magnetic field, $B_P$ is at an angle of less than $60^{\circ}$ with the disk surface \citep{Blandford1982}.  On the other hand, for weak $B_P$, the differential rotation of the disk can rapidly build up the toroidal magnetic field, potentially resulting in a magnetic pressure driven wind \citep[e.g.,][]{Uchida1985,Shibata1986,Stone1994,Contopoulos1995}.  Most likely, AGN winds are not driven by a single mechanism, but rather by some combination of all of them.  \citet{deKool1995} provide an example of this where gas is launched from the accretion disk and confined by a magnetic field, and then driven as a wind by UV radiation originating from the inner disk.

We emphasize that our approach is that of considering absorption features (NALs in particular) in the aggregate in contrast to investigating individual sources \citep[e.g.,][]{Richards2002,Moe2009,Hamann2011,Leighly2018} and we have not explicitly considered the relationship between optical/UV NALs and warm absorbers or ultra-fast outflows \citep{Reynolds1997,Tombesi2013}.  The goal of this current work is not so much to provide specific tests of the models discussed above as much as it is to explore the relationship of the prevalence of NALs (including AALs) to optical- and radio-luminosity and orientation (as potentially indicated by radio spectral index), which will in turn help to provide empirical data needed to construct tests of these different models for quasar outflows.

This paper is divided into the following sections.  Section~\ref{sec:data} describes the optical, radio, and \ion{C}{IV} NAL catalogs used, as well as any calculations for derived quantities.  In Section~\ref{sec:beta_NALs} we plot the velocity distributions of the \ion{C}{IV} NALs as a function of various quasar properties and take note of any observed trends.  We discuss the implications of the trends observed in Section~\ref{sec:discussion}.  Finally, we give our conclusions drawn from this work in Section~\ref{sec:conclusions}.  For this paper we assume a cosmology with the {\em Planck} 2015 cosmological parameters from \citet{Planck2015}.

\section{Data} \label{sec:data}

The data used for our analysis consists of two catalogs of absorbers drawn from two distinct samples of quasars from the Sloan Digital Sky Survey (SDSS; \citealt{York2000}) and radio data as described in detail in the following sections.

\subsection{Optical Data} \label{sec:opt_intro}
The quasars from which we draw the absorption-line sample are all objects that appear in one of two hand-vetted quasar catalogs produced by the SDSS project.  SDSS is a photometric and spectroscopic survey located at Apache Point Observatory.  The first phase of the survey, SDSS-I, ran from 2000-2005 and covered 8,000 square degrees of the sky in five optical bandpasses.  The second phase, SDSS-II, ran from 2005-2008 and extended the observation parameters of the survey with SEGUE and the Sloan Supernova Survey.  The spectroscopic survey in SDSS-I/II obtained medium resolution ($R \sim 2000$) spectra covering a wavelength range from \SIrange{3800}{9200}{\angstrom} and the final data set was released with the 7th Data Release \citep[DR7;][]{DR72009}.  The final SDSS I/II quasar catalog is described by \citet{DR7Q2010} and includes over 100,000 spectroscopically confirmed quasars, with spectra of sufficient quality to enable absorption line investigations.  The third phase of SDSS, SDSS-III, ran from 2008-2014, and is comprised of four separate surveys; here we considered the results up to Data Release 12 \citep[DR12;][]{DR122015}, specifically the quasar spectra from the Baryon Oscillation Spectroscopic Survey \citep[BOSS;][]{Dawson2013}.  The spectra provided by BOSS have a resolution of $1300<R<3000$ spanning a wavelength range of \SIrange{3600}{10400}{\angstrom} \citep{DR122015}.  In addition to the DR7 quasar catalog, our analysis also includes the 297,301 quasars included in the DR12 quasar catalog \citep{DR12Q2017}.  The selection algorithms for these SDSS quasar samples is far from uniform, but our analysis should not be significantly biased by the selection function and we have performed tests to ensure that our results are not sensitive to it.

Instead of taking the absolute magnitudes included in the SDSS catalogs, we used preliminary versions of the improved systemic quasar redshift measurements from Allen \& Hewett (2019, in prep., hereafter AH19) to calculate absolute magnitudes from the measured apparent $i$-band magnitudes.  Absolute magnitudes for our quasar sample are calculated according to: $M_i=m_i-A_i-DM-K(z)$, where $M_i$ is the $i$-band absolute magnitude, $m_i$ is the apparent $i$-band magnitude measured in SDSS DR7 or DR12, $A_i$ is the $i$-band galactic extinction based on the maps of \citet{Schlegel1998}, $DM$ is the distance modulus, and $K(z)$ is the K-correction from Table 4 of \citet{Richards2006b}.  On average the use of AH19 redshifts instead of SDSS pipeline redshifts changes $M_i$ by only 0.009.
  
An important spectral property when investigating NALs is the presence of BAL features, and connections (or lack thereof) between the two types of absorption systems may shed some light onto their origin.  To provide the most robust sample of NALs, quasars hosting BAL troughs have been removed from the sample: the DR7 quasars using the BAL list from \citet{Allen2011} and the DR12 quasars using the BAL flags from the catalog itself.  As such, this work will not compare NALs and BALs, but for completeness the relevant BAL properties are discussed here.

Finding the \citet{Weymann1991} definition of the ``BALnicity" index to be overly restrictive, \citet{Hall2002} created an ``intrinsic absorption index" (AI).  While objects that passed the classical BI criteria were unambiguously considered to be BALQSOs, there were many objects with C{\textsc{iv}} troughs that were classified as non-BALs by this definition, but nevertheless had many of the same characteristics of classical BAL troughs.  This included objects with troughs close to the systemic redshift, or troughs with significant contiguous widths despite being smaller than the BI definition allowed.  Much of the reasoning for the restrictive definition of BI was due to the resolution limits of the spectra at the time, and thus \citet{Weymann1991} wanted to clearly remove any false BAL classifications.  However, with the improved spectral resolution of SDSS, it became clear that there were intrinsic ``broad" absorption features that were being excluded from being considered a BAL.  Conversely, \citet{Knigge2008} point out that there is a bimodality in the AI distribution, in which only one mode is definitively related to the BAL phenomenon.  Thus, AI, while being more inclusive, may overestimate the fraction of objects with troughs due to outflows.  Ultimately, using BI is incomplete in its classification of BALQSOs, while using AI has the danger of including too many false positives.

\subsection{Radio Data}

As we are trying to factor out the contribution of intervening NALs from those are are intrinsic, radio data plays a key role in our analysis since the radio properties of a quasar are intrinsic to the quasar itself, and are unrelated to intervening galaxies.  As such, we investigate NALs as a function of radio luminosity and the relative strength of radio relative to the optical.  We also explore NALs as a function of the radio spectral index, which is thought to be correlated with quasar orientation \citep[e.g.,][]{Orr1982}.

Our primary source of radio brightness comes from the final Faint Images of the Radio Sky at Twenty centimeters (FIRST) catalog \citep{Helfand2015}.   FIRST used the Karl G. Jansky Very Large Array (VLA) in its B configuration to observe $>\SI[group-separator = {,}]{10000}{deg^2}$ of the night sky with an angular resolution of \ang[angle-symbol-over-decimal]{;;5.4} at \SI{1.4}{\GHz} (\SI{20}{\cm}) \citep{Becker1995, Helfand2015}.  Observing objects brighter than 1\,mJy, the final FIRST catalog contains more than 940,000 sources.  The area covered by FIRST is nearly the same area covered by SDSS \citep{York2000, Helfand2015}, allowing us to draw conclusions on the optical and radio properties of the many thousands of quasars observed by both surveys.  The data from FIRST is used to determine the radio-loudness (ratio of radio to optical flux; see Section~\ref{sec:alpha_r_intro}) and the radio luminosity.  The radio luminosities ($L_{1.4\mathrm{GHz}}$) are computed using the integrated flux measurements of FIRST, and then following Equation~3 in \citet{Kratzer2015} with z=2 K-corrections as in the manner of \citet{Richards2006b}:
\begin{equation} \label{eq:L_rad}
\frac{L_{\rm rad}}{4\pi\,{D}^2} = f_{\rm int}\,10^{-23}\,\frac{(1+2)^{\alpha_{\rm rad}}}{(1+z)^{(1+\alpha_{\rm rad})}},
\end{equation}
where $L_{\rm rad}$ is the radio luminosity measured in \si{\erg\per\s\per\Hz}, $D$ is the luminosity distance in \si{\cm}, $f_{\rm int}$ is the integrated flux in \si{\jansky}, $z$ is the quasar's systemic redshift, and $\alpha_{\rm rad}$ is the radio spectral index.  We again make use of the improved redshift measurements of AH19 in both $z$, directly in the above equation, and the calculation of luminosity distance.  Finally, for the radio spectral index, we assume $\alpha_{\rm rad}=-0.5$ as it is the traditional division between steep- and flat-spectrum objects.

In addition to FIRST, we also make use of the Green Bank \SI{6}{\cm} (GB6) survey \citep{Gregory1996}, which contains more than 75,000 sources brighter than \SI{25}{\milli\jansky}.  GB6 observed in the declination band $\ang{0}<\delta<\ang{75}$ with a beam size of $\ang[angle-symbol-over-decimal]{;3.6;}\times\ang[angle-symbol-over-decimal]{;3.4;}$ \citep{Gregory1996}.  The data from GB6 allows us to determine the radio spectral index between \SIlist{4.85; 1.4}{\GHz} by combining with the FIRST data (for a fraction of our sources).

\subsection{Radio Spectral Indices} \label{sec:alpha_r_intro}

Our analysis using the radio data will include investigations as a function of two radio-based spectral indices.  Here we define a spectral index, $\alpha$, where $f_{\nu} \propto \nu^{\alpha}$.  Thus, $\alpha$ can be determined if we have two flux density measurements at two known frequencies:
\begin{equation}
\alpha_r = \frac{\log(f_{\nu_1}/f_{\nu_2})}{\log(\nu_1/\nu_2)},
\end{equation}
where $\nu_2>\nu_1$.

The in-band radio spectral index, which we designate as $\alpha_{\rm r}$ is an important property to consider as it is thought to be correlated with orientation in jet-dominated quasars \citep[e.g.,][]{Orr1982,Padovani1992,Wills1995}.  Traditionally, the radio band has been used in a variety of ways to determine the orientation of the quasar with respect to the beaming of the radio jets along our line of sight.  The parameters measured include the core-to-lobe flux ratio, $R$ or $C$, the ratio of the core radio luminosity to optical luminosity, $R_V$ \citep{Wills1995,Barthel2000},  and the in-band radio spectral index, $\alpha_{\rm r}$; see \citet[][and references therein]{Richards2001a}.  Regardless of whether $\alpha_{\rm r}$ is a robust indicator of quasar orientation, it is at the very least an intrinsic property of quasars, and should not correlate with properties of intervening absorption systems. 

Most of our spectral index measurements are in the form of $\alpha^{1.4}_{4.85}$, where the \SI{1.4}{\GHz} data are integrated fluxes measured by FIRST and the \SI{4.85}{\GHz} measurements come from GB6.  However, the resolution of FIRST (\ang[angle-symbol-over-decimal]{;;5.4} beam) is much higher than that of GB6 ($\ang[angle-symbol-over-decimal]{;3.6;}\times\ang[angle-symbol-over-decimal]{;3.4;}$ beam).  To account for the lower resolution of GB6, we took all of the FIRST sources that would have been within the beam of GB6 and added their integrated fluxes together, creating a combined FIRST flux measurement with an effective beam comparable to GB6.  As the FIRST and GB6 data are not taken simultaneously, the spectral indices are subject to systematic errors.  In particular, we might expect steep radio spectrum quasars to be less subject to radio variability than (presumably more face-on) flat radio-spectrum quasars ($\alpha>-0.5$).

The VLA Sky Survey (VLASS; \citealt{Murphy2015}) will solve the problem of non-simultaneity of the radio bandpasses by taking observations that simultaneously cover \SIrange[range-phrase=--]{2}{4}{\GHz} (S-band) over nearly \SI[group-separator = {,}]{30000}{\deg\squared} of sky, allowing determination of the spectral index (at $\sim$\SI{3}{\GHz}) from a single data set.  The (completed) Caltech NRAO SDSS Stripe 82 Survey (CNSS; \citealt{Mooley2016}) is very similar in structure to (the ongoing) VLASS and can be used as a preliminary indicator of  what we can expect from VLASS and provides a check on radio spectral indices determined from non-simultaneous observations in different bands.  As our analysis is differential in nature, the absolute accuracy of $\alpha_{\rm rad}$ due to non-simultaneous observations is not the limiting factor in our investigation.  We are not currently making use of the data from either of these sources; however, such data would support future analysis that requires greater certainty in the measurement of individual values of $\alpha_{\rm rad}$. 

Another type of spectral index we consider is one of the ``radio loudness" measures of quasars.  ``Radio loudness" can be determined by the radio luminosity alone \citep[e.g.,][]{Peacock1986} or by the ratio of the radio to the optical luminosity \citep[e.g.,][]{Kellerman1989}, which is just the cross-band spectral index, $\alpha_{ro}$.  As there is disagreement on whether the radio luminosity or the radio-to-optical spectral index is the best measure of radio loudness \citep{Goldschmidt1999,Ivezic2002,Balokovic2012,Kratzer2015}, we will explore whether there are significant trends in NAL properties as a function of {\em both} measures.

Radio loudness can be determined from the ratio of radio-to-optical emission given by:
\begin{equation}
R=\frac{f_{\SI{6}{\cm}}}{f_{\SI{4400}{\angstrom}}}
\end{equation}
where $f_{6\,\si{\cm}}$ is the \SI{6}{\cm} (\SI{4.85}{\GHz}) radio flux, $f_{\SI{4400}{\angstrom}}$ is the \SI{4400}{\angstrom} optical flux.  Typically, sources with $\log{R}>1$ are considered radio loud.  Although $R$ and $\log{R}$ are more commonly used in the literature and familiar to radio astronomers, we will follow \citet{Kratzer2015} in using the terminology of spectral indices, with $\alpha_{ro}$ as our defining parameter of the radio loudness of a quasar.  The radio-optical spectral index ($\alpha_{\mathrm{ro}}$) is computed as follows:
\begin{equation}
\alpha_{\mathrm{ro}} = \frac{\log(L_{\SI{20}{\cm}}/L_{\SI{2500}{\angstrom}})}{\log(\SI{20}{\cm}/\SI{2500}{\angstrom})},
\end{equation}
where $L_{\SI{20}{\cm}}$ is the FIRST (\SI{1.4}{GHz}) radio luminosity (see Equation~\ref{eq:L_rad}), and $L_{\SI{2500}{\angstrom}}$ is the \SI{2500}{\angstrom} optical luminosity, calculated from $M_i(z=2)$ following Equation~4 in \citet{Richards2006b}.   $\alpha_{\rm ro}=-0.2$ is roughly equivalent to $\log R = 1$ and a very radio-loud quasar with $\log R = 3$ would have $\alpha_{\rm ro}=-0.6$.

\subsection{Absorption Line Data} \label{sec:abs_data_intro}
Absorption line catalogs from both the DR7 and DR12 quasars were kindly provided by P. Hewett (private communication, 2018).  The DR7 absorber catalog is described in \citet{Bowler2014}, but briefly can be summarized as follows.  There are 13,759 quasars containing a total of 39,511 absorbers with rest-frame equivalent widths ($\rm EW_{rest}$ or $\rm EW_r$) ranging from \SIrange{0.034}{76.55}{\angstrom}.  Of these roughly 40k absorbers, 17,566 are \ion{C}{IV} absorbers, the other 22,945 are \ion{Mg}{II} absorbers.  The DR12 absorber catalog was compiled using a scheme as described in Section~2 of \citet{Wild2006}, albeit with \ion{C}{IV} doublet components rather than \ion{Mg}{II} or \ion{Ca}{II}, and contains 58,757 quasars with 105,234 \ion{C}{IV} absorbers, with rest-frame equivalent widths spanning \SIrange{0.04}{5.51}{\angstrom}.  Based on analysis similar to that shown in Figure~4 of \citet{Bowler2014}, we estimate that 1.5\% of the DR12 systems identified as \ion{C}{IV} are instead misidentified lines of \ion{Si}{IV}, \ion{Al}{II}, \ion{Fe}{II}, or \ion{Si}{II}. For any quasars with absorption data in both DR7 and DR12 catalogs, only the DR12 absorbers were included in this work, reducing the DR7 sample to 2,923 \ion{C}{IV} absorbers. 

\begin{figure*}
	\centering
	\begin{subfigure}{0.45\textwidth}
		\includegraphics[width=\textwidth]{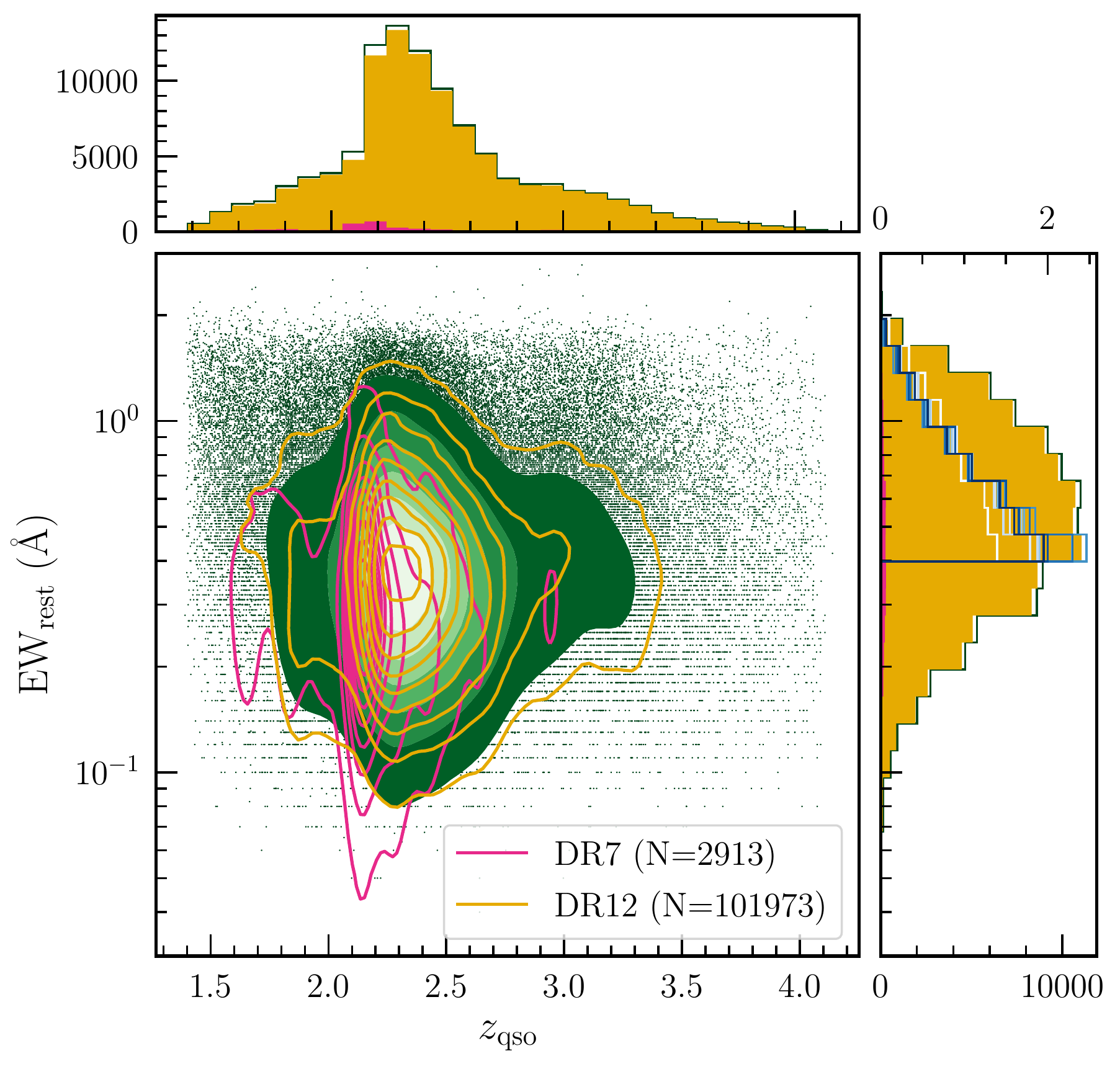}
	\end{subfigure}
	\hfill
	\begin{subfigure}{0.45\textwidth}
		\includegraphics[width=\textwidth]{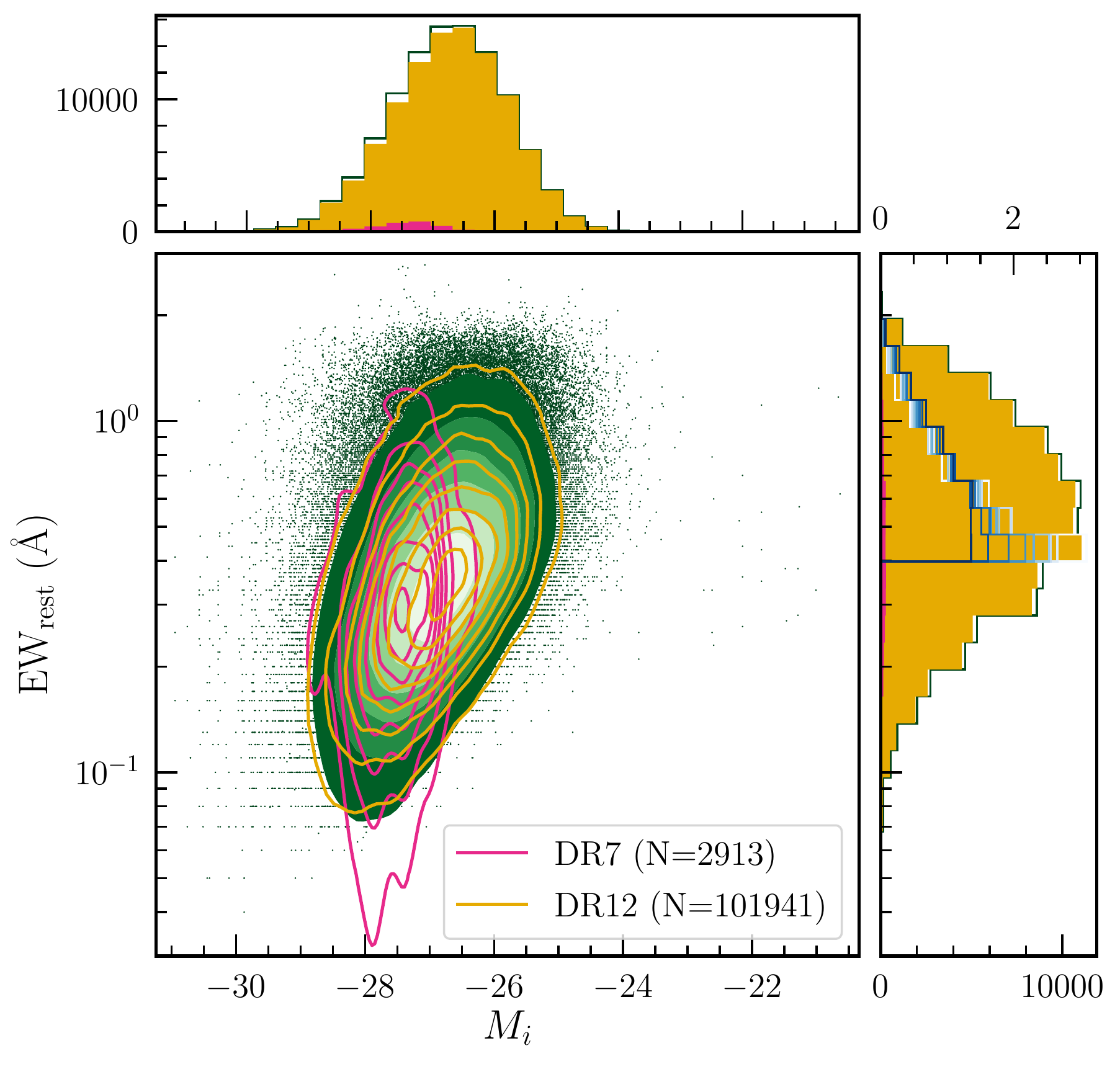}
	\end{subfigure}
	\caption{(\emph{Left}) Redshift and rest-frame equivalent width ($\rm EW_r$) distribution of the absorber sample. The distribution suggests that our sample is probing roughly the same rest EW distribution across the full redshift space sampled.  That there is a lack of redshift dependence in $\rm EW_r$ is confirmed by the blue-shaded histograms in the right sub-panel, which shows slices in redshift (for $\rm EW_r>0.4$ where the sample is most complete), normalized by the total number of absorbers in each slice.  (\emph{Right}) \emph{i}-band absolute magnitude and rest-frame equivalent width distribution of the absorber sample.  There is an apparent trend between $\rm EW_r$ and $M_i$ as it is easier to measure smaller $\rm EW_r$ absorbers in more luminous quasars (causing ``missing" absorbers in the lower right) and due to an overall smaller number of luminous quasars (causing ``missing" absorbers in the upper left).  The blue-shaded histograms in the right sub-panel, showing slices in $M_i$ (again, above $\rm EW_r>0.4$), illustrate that the apparent dependence of $\rm EW_r$ with $M_i$ is due to selection effects. (\emph{Both}) The main panels are kernel density estimates of the 2D parameter spaces.  The green shades represent the full data sample, while the pink and yellow contours show the portions of the data from DR7 and DR12, respectively.   The top horizontal plots and right vertical plots show the full distributions of their respective properties (tops: $z_{\rm qso}$ and $M_i$; rights [bottom axis]; $\rm EW_r$).  The blue histograms in the vertical plots are density plots of the $\rm EW_r$ distributions of 9 equally sized slices of $z_{\rm qso}$ and $M_i$, respectively.  These histograms show the relative uniformity of $\rm EW_r$ in the range \SIrange{0.4}{1.0}{\angstrom} with respect to redshift and absolute magnitude.}
	\label{fig:EW_dist}
\end{figure*}

The DR12 sample shares many of the same characteristics as the DR7 sample from \citet{Bowler2014}, which can be summarized as follows.  The absorber sample is effectively complete for equivalent widths in the range \SIrange{0.4}{1.0}{\angstrom} (see Figure~\ref{fig:EW_dist}).  Above \SI{1.0}{\angstrom}, the sample rapidly becomes incomplete due to a detection threshold imposed to ensure the individual lines in the doublet are resolved.  Below \SI{0.4}{\angstrom} the sample is incomplete from the dependence on high S/N in quasar spectra necessary for the detection of low $\rm EW_r$ absorbers.  The differential nature of this work is largely indifferent to the incompleteness across $\rm EW_r$, but benefits from maximizing the number the \ion{C}{IV} absorbers that are included.  Therefore, as in \citet{Bowler2014}, we have limited the equivalent widths of the full absorber sample to have the interval $0.1\leq\rm EW_r<\SI{3.0}{\angstrom}$ (the upper limit removes only 12 absorbers with excessively large $\rm EW_r$).  Finally, we removed DR12 quasar spectra with low S/N ($<5$), altogether reducing the absorber count to 2,895 and 101,700 in the DR7 and DR12 samples, respectively.

Both of these catalogs benefit significantly from improved systemic redshifts for the quasars, as described by AH19, which builds upon the work of \citet[][hereafter HW10]{Hewett2010}.\footnote{Note that the improvements in the redshifts from \citet{Hewett2010} over the SDSS pipeline redshifts are far more significant than the improvements in the redshifts from HW10 to AH19.}  Accurate systemic redshifts are important as emission-line redshifts are subject to a higher level of uncertainty (the FWHM of the broad emission line being thousands of \si{\km\per\s}) than the redshifts of the absorbers and small changes in the systemic redshift have a significant effect on our interpretation of either 1) the physical distance between the quasar and absorber (if intervening) or 2) the outflow (or inflow) velocity of the absorber (if intrinsic).

Our full data sample consists of 104,595 absorbers from 57,248 quasars.  The QSO rest-frame equivalent width distribution as a function of the redshift distribution and as a function of $i$-band absolute magnitude can be seen in Figure~\ref{fig:EW_dist}\footnote{The slightly discrepant number of absorbers in the DR12 sample in the right plot are from a very small number of quasars with missing apparent magnitudes.}.  The upper and lower limits of the rest EW distributions are relatively uniform across the full redshift range sampled. Figure~\ref{fig:EW_dist} would seem to suggest that the equivalent widths and absolute magnitudes are (anti-)correlated.  However, that this effect is not real can be understood as follows:  1)  it is more difficult to find low $\rm EW_{r}$ absorbers in low-luminosity objects due to the low S/N in their spectra,  and 2)  the apparent dearth of high $\rm EW_{r}$ absorbers from bright sources is just an artifact of the small numbers of such objects. Truncating the sample  where $\rm EW_{r}$ becomes incomplete and then normalizing by the total (blue histograms in the right-hand vertical plot of Figure~\ref{fig:EW_dist}), reveals no trend between $\rm EW_{r}$ and $M_i$.

\begin{figure}
	\centering
	\includegraphics[width=0.45\textwidth]{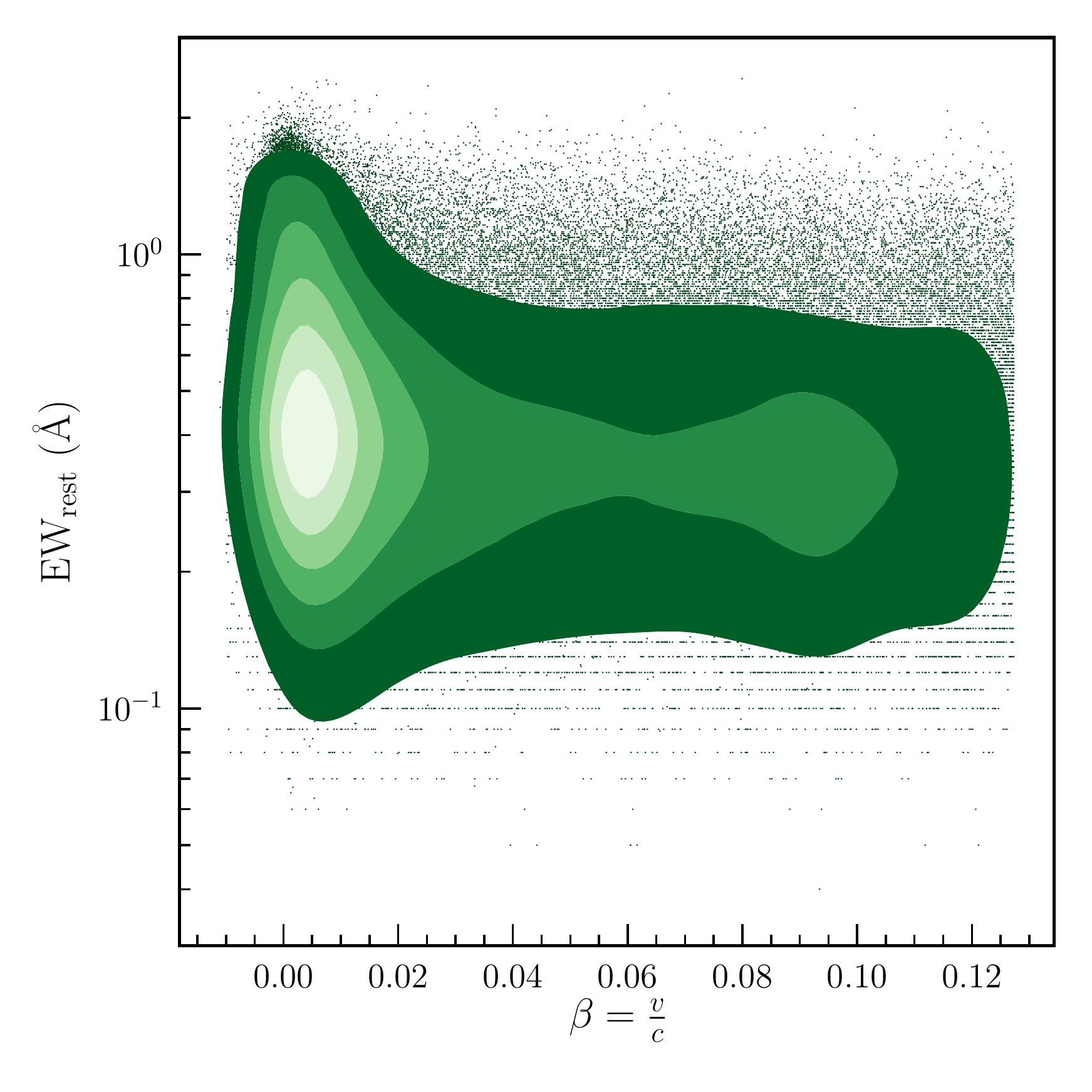}
	\caption{Rest-frame equivalent width ($\rm EW_r$) versus absorber velocity ($\beta$).  The weakest absorber equivalent widths appear to be relatively equally distributed across velocity space.  There is a small excess of weak absorbers at low velocity, but nowhere near the excess of {\em strong} absorbers at low velocity and there are weak absorbers at all velocities investigated.  This distribution suggests that there is no strong bias towards only observing the weakest absorbers in the high S/N regime near the BEL peak, whereas the excess of the strongest $\rm EW_r$'s appearing at low velocity most likely has a physical origin.}
	\label{fig:EWr_beta}
\end{figure}

Another important consideration is that the signal-to-noise of the spectrum is higher within the emission line profile.  Thus we might expect that, if noisier spectra were to contain narrow absorption lines, then they would be preferentially found within the emission line profile.  If true, such a finding would bias our conclusions when comparing the low-velocity absorbers (those found on top of the emission lines) to the high-velocity absorbers, since we might not sample the full population of high-velocity absorbers.  However, a plot of absorber rest-frame equivalent width versus velocity (Figure~\ref{fig:EWr_beta}) demonstrates that there is little bias in observing the weakest absorption at either high or low velocity. (And, moreover, our analysis is limited to absorbers with $\rm EW_r>\SI{0.1}{\angstrom}$.) On the contrary, the low-velocity absorption exhibits an excess of the {\em strongest} absorption lines compared to the high-velocity absorption.  As these lines should be as easily observed at high velocity as at low velocity, their preference for appearing at low velocities is most likely physical in nature as we will discuss in Section~\ref{sec:discussion}.

\section{Intrinsic versus Intervening NALs} \label{sec:beta_NALs}

With the right data (e.g., sufficiently high spectral resolution) it is possible to determine whether a particular NAL is more likely to be intrinsic or intervening (e.g., using partial covering, see \citealt{Hamann2011}).  However, even without such data, we can gain significant insight into quasar outflows by exploring the distribution of NALs in bulk.  In particular we can use the fact that intervening NALs should be independent of quasar-intrinsic properties to learn about the properties of intrinsic NALs---even without identifying which particular NALs are intrinsic.

In this section we first investigate the consequences of the (preliminary) improved systemic quasar redshifts to be published in AH19 and their effect on the absorber velocity distributions.  We then look into the correlations of the velocity distributions as functions of various radio properties, including radio detection, radio morphology, radio loudness, radio luminosity, and radio spectral index.  We then shift to the optical by discussing the correlation of absolute magnitude with absorber velocity, and also how normalizing the data by absolute magnitude may be important for our sample.  Finally, we compare our results to those in \citet{Richards2001b}, which is one of the driving motivators for this work.

\subsection{Insights from Systemic Redshifts} \label{sec:beta_dist_z}
To reiterate, by virtue of being narrow lines, the redshifts of NALs (whether they are purely cosmological or also affected by a peculiar velocity component) are determined far more accurately than the redshifts of the broad emission lines.  It is, however, the BELs that determine a quasar's systemic redshift.  Uncertainty in the systemic redshift of a quasar leads to a systematic error in the relative velocities of NALs that are truly intrinsic.  This error can significantly affect the interpretation of the nature of these systems. 

\begin{figure*}
	\centering
	\begin{subfigure}{0.45\textwidth}
		\includegraphics[width=\textwidth]{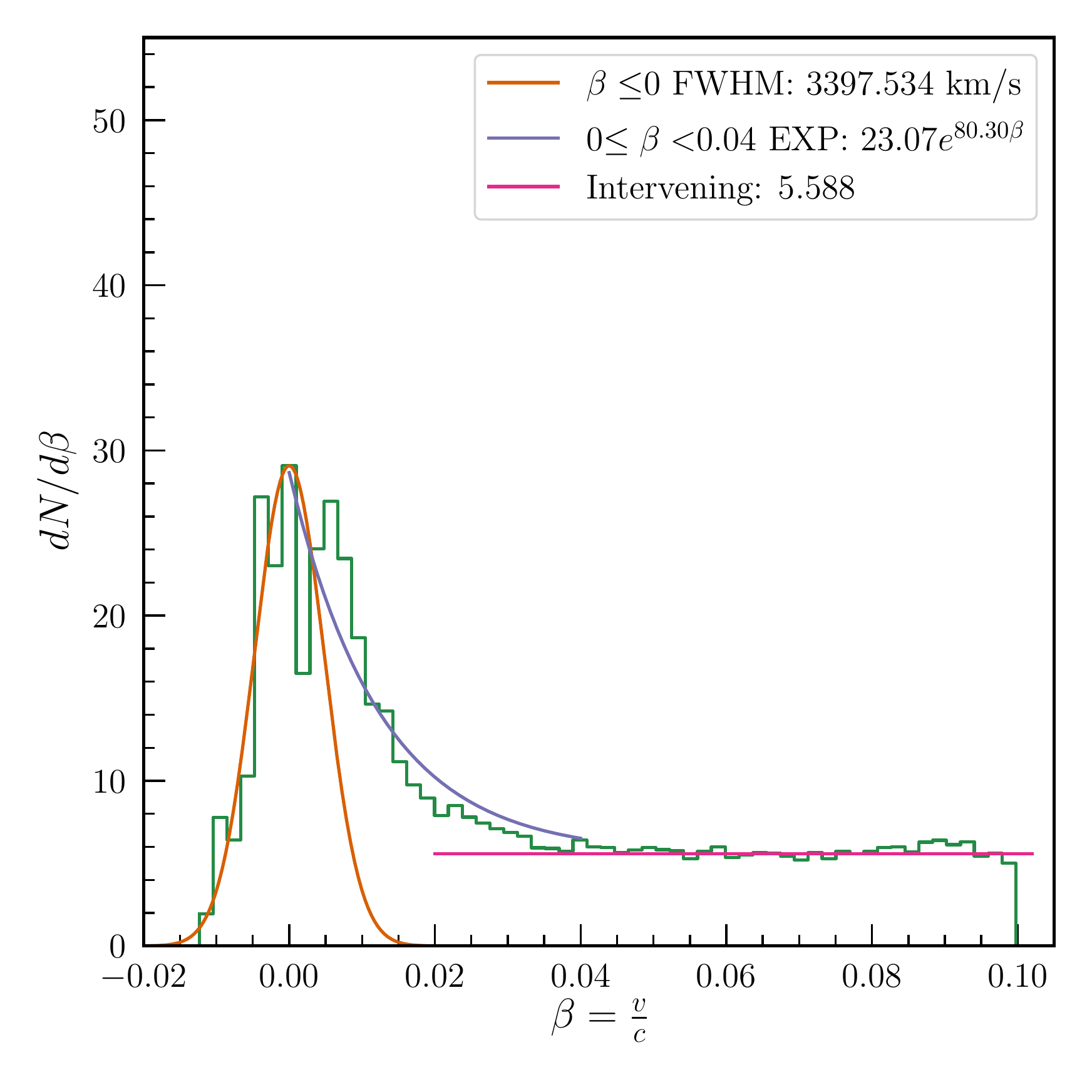}
	\end{subfigure}
	\hfill
	\begin{subfigure}{0.45\textwidth}
		\includegraphics[width=\textwidth]{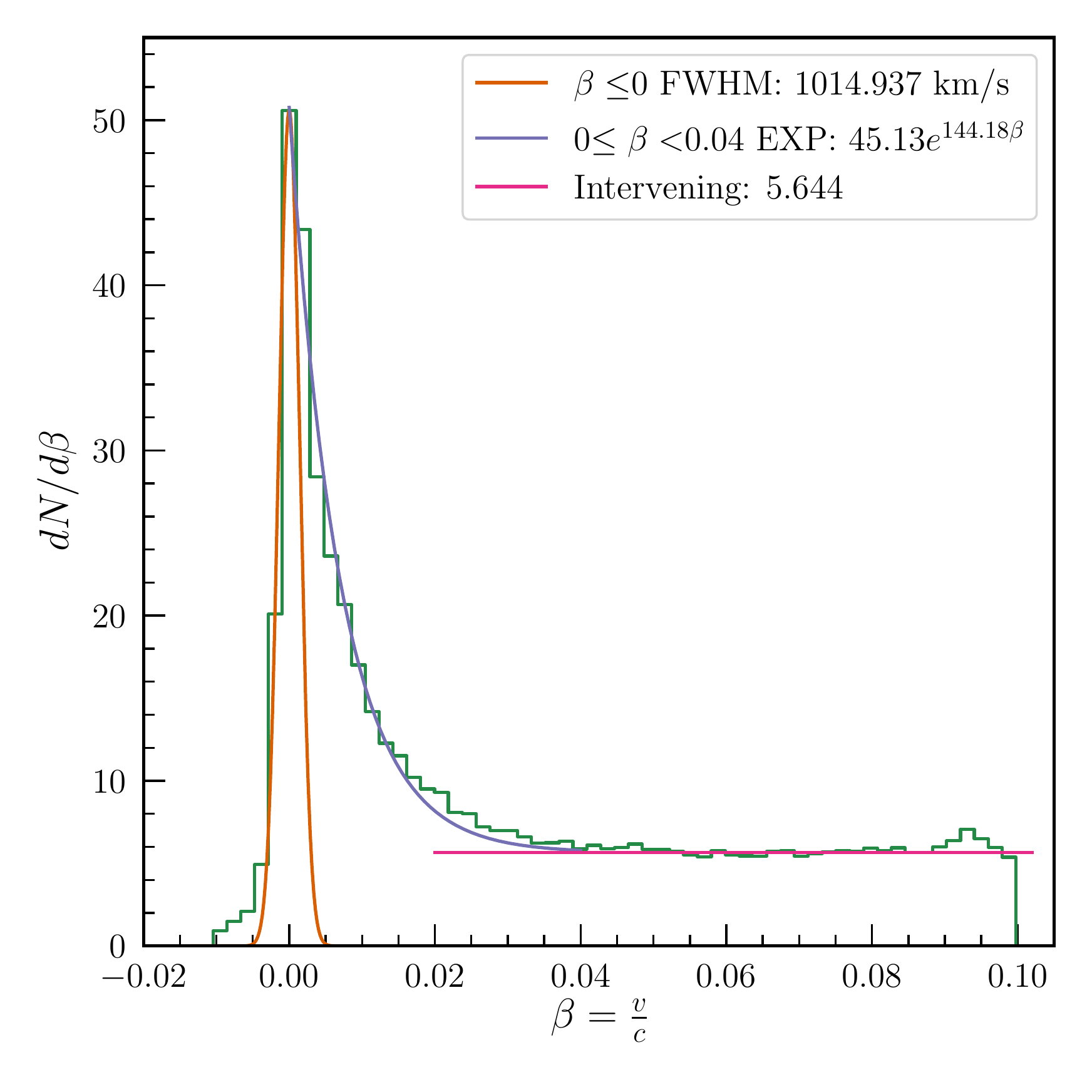}
	\end{subfigure}
	\caption{Velocity distribution of \ion{C}{IV} absorption.  The velocities of the absorbers are calculated using the quasar redshifts from the SDSS pipeline (\emph{left}) and the quasar redshifts computed from the work of AH19 (\emph{right}). Both figures contain the same total number of absorbers and each shows all three components (to guide the eye, not a fit to the total distribution) of the velocity distribution.  \emph{Orange}: Low-velocity intrinsic absorbers centered at \SI{0}{\km\per\s} (Gaussian), \emph{Purple}: High-velocity absorber tail (exponential), \emph{Pink}: Intervening systems, presumably galaxies (average).  With improved redshift measurements from AH19, we see a smaller FWHM in the Gaussian and that there is a smaller total contribution to the AALs from the clustering component than would be predicted using SDSS pipeline redshifts.  The small bump above the average at $\beta\sim0.094$ (roughly \SI{6}{\percent} excess in those bins) occurs as the result of  residual unmasked contamination of the \ion{C}{IV} sample by \ion{Si}{IV}.}
	\label{fig:zimprov}
\end{figure*}

In Figure~\ref{fig:zimprov} we show in the left-hand panel the velocity distribution of \ion{C}{IV} NALs (assuming that all are intrinsic) using the systemic redshift as derived from the SDSS pipeline ({\tt Z\_PIPE} from \citealt{DR12Q2017}, see Section~4.5 for a discussion on the accuracies of the various redshift measurements included in the DR12 quasar catalog).  The distribution looks very much like that of \citet{Weymann1979}, who found that the velocity distribution of NALs has three distinct components: 1) a very high velocity component that is roughly uniformly distributed and thus more likely to represent intervening material that is unassociated with the quasar; (2) a Gaussian peak centered at $v=0$ and extending a few thousand \si{\km\per\s} to either side, presumably caused by virialized material clustered around the quasar itself; and (3) a high velocity asymmetrical tail on the positive side of the peak possibly associated with material ejected from the QSO \citep{Foltz1986,Weymann1979}.

In the right-hand panel of Figure~\ref{fig:zimprov} we instead show the velocity distribution of the same NALs systems, but now using systemic redshifts as derived from AH19.  What is immediately obvious is the much narrower distribution of the ``Gaussian" component.  The fact that the new quasar redshifts were derived completely independent of any information about the NALs and the fact that it is very hard to make a Gaussian distribution {\em narrower} as a result of making a systematic error, provides further support that the systemic quasar redshifts are more accurate.  This result has the effect of increasing the number of NALs that are contained in the ``outflowing" component and reducing the contribution of those NALs that are most consistent with the ``virialized" component (deduced from estimates of the areas under the curves).  Little change is seen in the ``intervening" component.  Both panels of Figure~\ref{fig:zimprov} are shown on the same scale in order to highlight the differences that results from using more accurate systemic redshifts for the quasars.  Figure~15 of \citet{Hewett2010} shows a directly analogous comparison as in Figure~\ref{fig:zimprov} of this paper, demonstrating improved systemic redshifts for a sample of SDSS DR6 quasars.

Thus, even without any information about individual NALs, we can determine that the vast majority of NALs with relative velocities between $\sim \SI{1500}{\km\per\s}$ and $\sim \SI{6000}{\km\per\s}$ represent material that is outflowing from the host quasar and thus should perhaps be treated as an extension of the BAL distribution \citep[e.g.,][]{Nestor2008,Wild2008,Bowler2014}.  This is true even though $\Delta v/c = \SI{3000}{\km\per\s}$ represents 44 comoving Mpc, and that these systems might have, a priori, been expected to be intervening.  \citet{Bowler2014} provide further evidence for some high-velocity NALs being intrinsic as line-locking between systems seen in the same quasar spectrum could provide evidence that a fraction of the NALs are driven by radiation pressure.

It is less clear what the ``virialized" component represents.  Those systems could be the result of material (clouds, dwarf galaxies, etc.) gravitationally bound to the quasar system \citep[e.g.,][]{Richards2001b,Nestor2008,Wild2008}.  While there may still be some residual errors in the quasar systemic redshift measurements, any further improvements are unlikely to move more absorbers from this component to the outflowing component.  This can be seen by analyzing the velocity distribution using the quasar redshifts of \citet{Hewett2010}, which represent an intermediate improvement between the pipeline redshifts and the AH19 redshifts.  The width of the Gaussian component tightens each time redshifts improve, but rough estimates of the areas under the curve do not change between \citet{Hewett2010} and AH19.  In other words, while the shape of the velocity distribution changes and suggests improved redshift measures, the actual number of absorbers in the ``virialized" component remains the same, thus their origin is also unchanged.  Further improvements to the systemic redshifts might incrementally narrow the Gaussian component, but will never eliminate it completely.

In the right plot of Figure~\ref{fig:zimprov} a small bump above the mean level (roughly \SI{6}{\percent} excess in those bins) appears at $\beta\sim0.094$, which is not seen in the left plot.  The inaccuracies of the SDSS pipeline systemic redshifts spread out the apparent velocities of the aborbers, effectively hiding the excess.  However, this excess in the right panel reveals both that the emission line redshifts have been improved and that there is a small amount of residual contamination of the \ion{C}{IV} sample from unmasked \ion{Si}{IV} absorption.

\subsection{Insights from Radio Properties} \label{sec:beta_dist_rad}

Key insight to the nature of NALs comes from investigating them as a function of the radio properties of their host quasars.  Again, intervening systems should be immune to the details of the quasar radio emission, so any radio-dependent effects are indicative of a population of intrinsic NALs.  

Early work found significant differences in the velocity distribution of NALs (particularly AALs) between steep- and flat-spectrum radio sources \citep{Anderson1987, Foltz1986}, which is illustrated in Figure~31 of \citet{Richards2001b} and Figure~7 of \citet{Richards2001a}.  \citet{Richards2001b} further suggested that there may be differences between radio-loud and radio-quiet quasars and not just the radio spectral indices; see their Figure~28.

Thus, in a similar fashion, we separated our absorber sample into groups based on radio spectral index and radio-optical spectral index (as a measure of radio ``loudness", see Section~\ref{sec:alpha_r_intro}).  For each property we separated the {\em quasars} into two or three subgroups with an equal number of objects in each---in order to identify any trends with the radio properties.  We then investigated the properties of the absorbers in each group.  Since an individual quasar could have multiple absorbers observed in its spectrum (or none), this process led to unequal numbers of absorbers in each group (which itself is potentially interesting).  For nearly all of the normalized velocity distribution (${\rm d}N/{\rm d}\beta$) plots in the following sections we also included an inset plot where we restricted the bin sizes to match the components detailed in Section~\ref{sec:beta_dist_z} (i.e. low-velocity gaussian, high-velocity exponential tail, and very high-velocity intervening component), which allows a more a precise comparison to earlier work.  We further use the same maximum $y$-axis value in (nearly) all of these plots to better illustrate the changes in the low-velocity distribution of NALs with various properties.  Lastly, as the velocity distributions in the following sections are \emph{normalized}, the total number of absorbers that occupy each subgroup are included in the ${\rm d}N/{\rm d}\beta$ plots since those numbers cannot be deduced directly from the plots.

\subsubsection{Radio Loudness} \label{sec:radio_loudness}
\begin{figure*}
	\centering
	\begin{subfigure}{0.45\textwidth}
		\includegraphics[width=\textwidth]{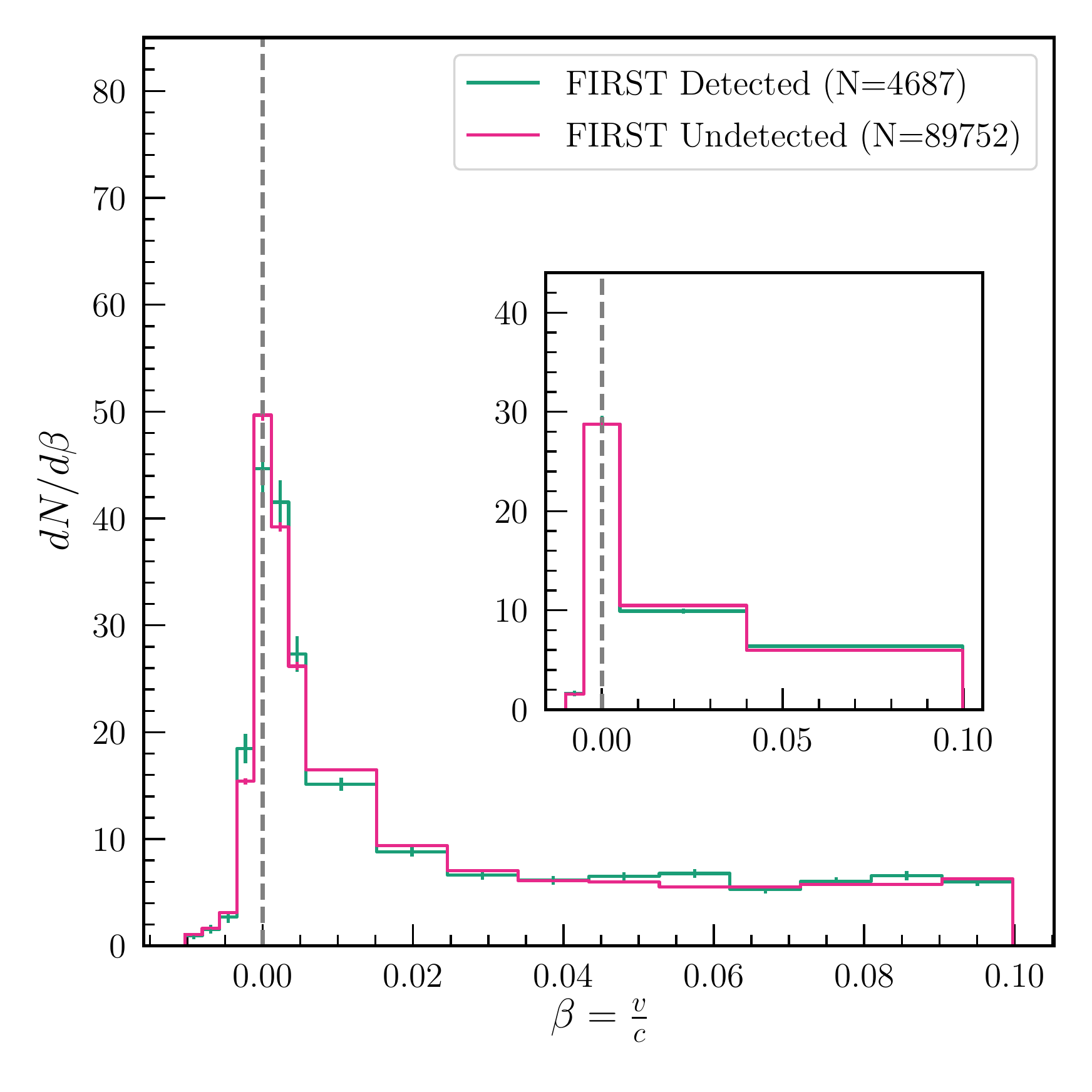}
	\end{subfigure}
	\hfill
	\begin{subfigure}{0.45\textwidth}
		\includegraphics[width=\textwidth]{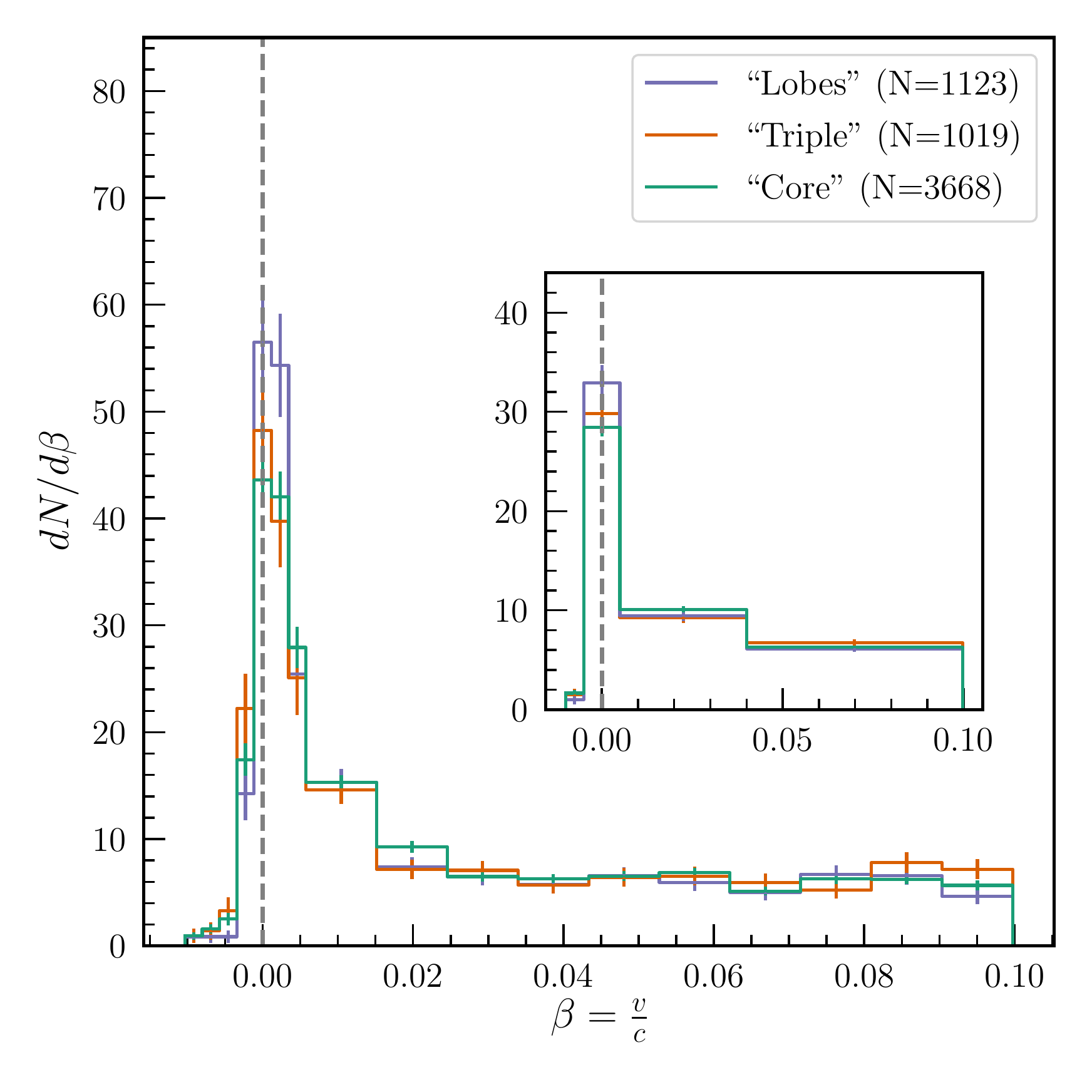}
	\end{subfigure}
	\caption{(\emph{Left}) Absorber velocity distributions (FIRST detected/undetected). This plot separates absorbers based on whether or not the host quasar was detected within the FIRST footprint.  We see no significant difference between the two groups, suggesting that the existence of radio emission (or lack thereof) does not directly affect the physics governing NALs. This is even more apparent from the inset plot, which shows the same data but with only 4 bins corresponding to the absorption components detailed in Section~\ref{sec:beta_dist_z}.  The grey dashed line is meant only to guide the eye to an absorption velocity of $v=0$.  (\emph{Right}) Absorber velocity distributions (extended radio emission). This plot separates absorbers into groups based on the host quasar's extended radio emission, as detailed in the text.  There is a relatively weak difference between these groups at low velocity, but no trend stands out at high velocities.}
	\label{fig:rad_det}
\end{figure*}

We begin examining the NAL distribution as a function of radio loudness by simply comparing those objects that are  FIRST radio detections and those that are not.  For the non-detection sample we made sure to include only those quasars that fall within the FIRST footprint, but lack a \SI{1.4}{\GHz} flux measurement.  These two samples are compared in the left panel of Figure~\ref{fig:rad_det}, where the similarity suggests that whether or not a quasar has any radio emission does not affect the physics behind the (narrow) absorber outflows: both radio-detected and non-detected quasars have similar numbers of AALs at low velocity, likely outflowing systems at moderate velocity, and likely intrinsic systems at high velocity.

We also investigated what affect the general radio morphology of the quasar would have on the NAL distribution.  In that regard we created two classes by matching FIRST to SDSS quasars.  If there was a single match within \ang{;;2}, the radio source was considered to be from the ``core", and if there was one or more matches within \ang{;;60}, the radio source(s) was considered to be from extended emission.  Once we completed this matching, we created three groups for this extended radio emission property: objects with both extended and core radio emission (``triple"), those with just extended radio emission (no core; ``lobes"), and those with only core emission (``core").  We plot the absorber velocity distribution for each of the three groups in Figure~\ref{fig:rad_det}.  At high velocities there is no significant difference between the three groups.  However, there is a relatively weak trend at low-velocities, with the ``lobes" group exhibiting the strongest associated absorption, followed by the ``triple" group, and the ``core" group showing (marginally) weaker associated absorption.

\begin{figure*}
	\centering
	\begin{subfigure}{0.45\textwidth}
		\includegraphics[width=\textwidth]{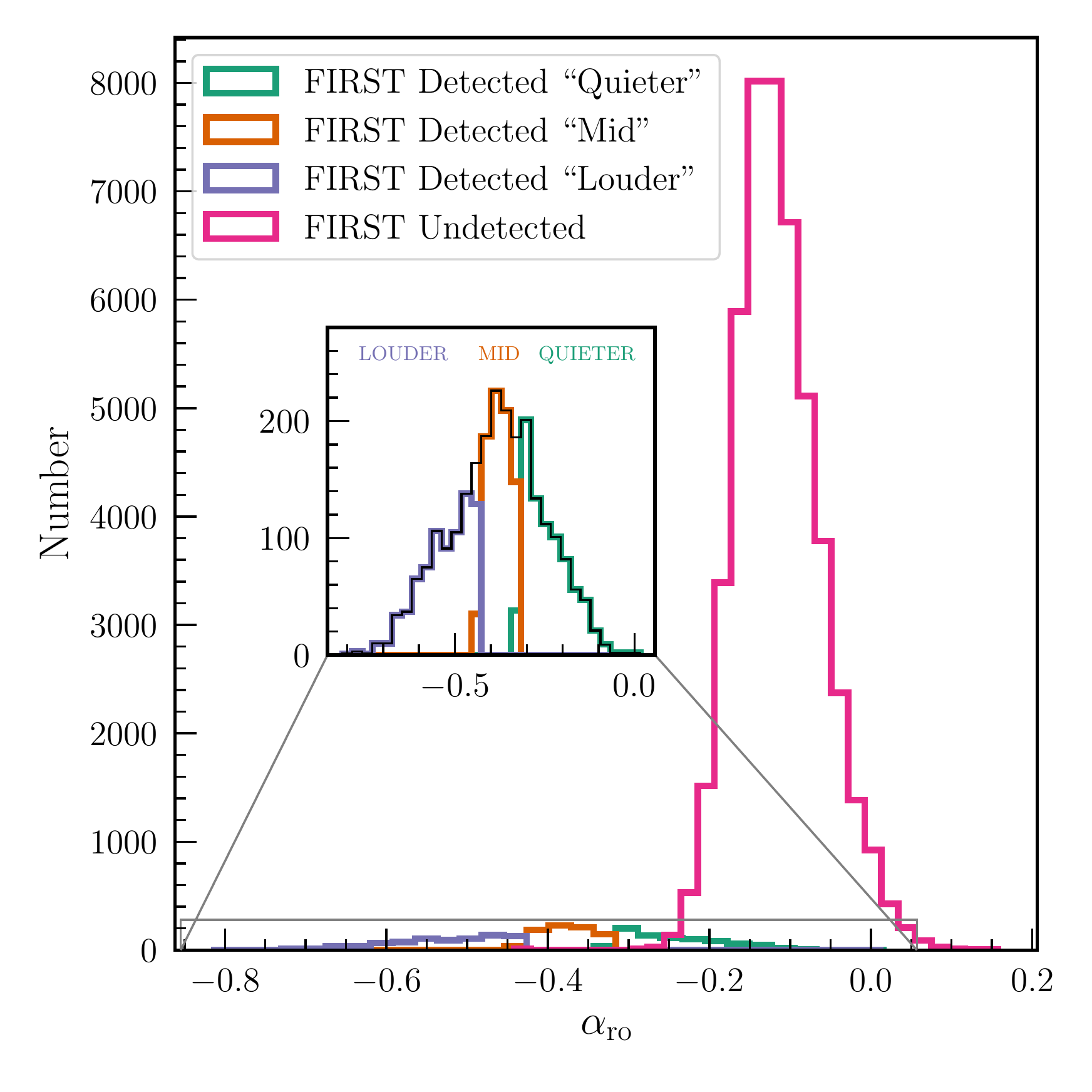}
	\end{subfigure}
	\hfill
	\begin{subfigure}{0.45\textwidth}
		\includegraphics[width=\textwidth]{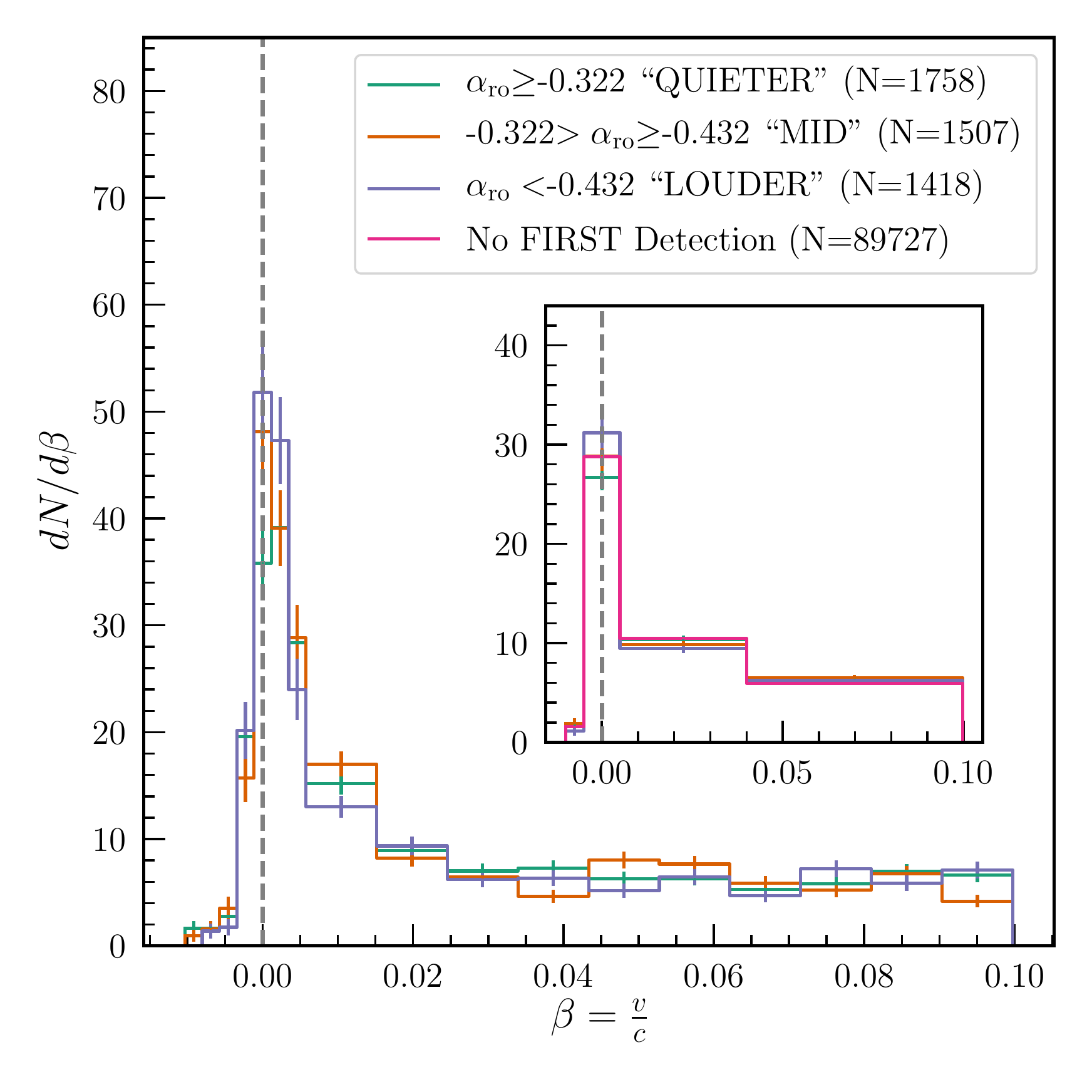}
	\end{subfigure}
	\caption{(\emph{Left}) Distribution of radio-optical spectral indices ($\alpha_{\mathrm{ro}}$; three groups) including those computed for radio non-detections assuming an observed flux of \SI{0.2}{\milli\jansky} (upper limit).  The solid-colored lines in the inset indicate the separation of the quasars with radio detections into three equal groups. (\emph{Right}) Absorber velocity distributions ($\alpha_{ro}$ spectral index groups along with the non-detection group).  At $v\sim\SI{0}{\km\per\s}$ there appears to be a weak excess of intrinsic absorption in the radio-loudest quasars relative to the radio-quietest (but still radio detected). At higher velocities there does not appear to be any specific trend.}
	\label{fig:alpha_ro_wNonRad}
\end{figure*}

In Figure~\ref{fig:alpha_ro_wNonRad} we investigate distribution of NALs (right panel) as a function of the radio loudness---as measured by $\alpha_{\rm ro}$ (see Section~\ref{sec:alpha_r_intro}), breaking the ``FIRST detected" radio sample (left plot of Figure~\ref{fig:rad_det}) into three equal bins of quasars (as shown in the left panel).  Recall that $\alpha_{\rm ro}=-0.2$ is roughly equivalent to the traditional loud/quiet dividing line.   We also include the velocity distribution of the absorbers from all of the quasars that were not detected by FIRST (this non-detection group is the same as in the left plot of Figure~\ref{fig:rad_det}).  As FIRST has a detection limit of \SI{1}{\milli\jansky}, we set the observed flux for all of the radio non-detected quasars to \SI{0.2}{\milli\jansky} (5 times lower than the FIRST detection limit).  This procedure provides upper limits for all of the non-detections, allowing us to make some general comparisons about their radio loudness and the velocity distribution of the NALs in these quasars, particularly in comparison to the three separate ``FIRST detected" groups.  As our divisions do not adhere to the traditional criteria of loud and quiet quasars, we instead refer to them as radio-``louder" and radio-``quieter".

The results (from the right panel) do not appear to agree with one of the findings of \citet{Richards2001b}, who found a small excess of high velocity absorbers in radio-quiet quasars.  In our data set the high velocity absorbers do not appear to prefer any particular radio loudness.  However, there is a small excess of the systems with $v\sim\SI{0}{\km\per\s}$ in the radio loudest quasars.  Again, we find that the radio non-detection group does not appear to differ from the mean of the radio-detected groups.

\begin{figure*}
	\centering
	\begin{subfigure}{0.45\textwidth}
		\includegraphics[width=\textwidth]{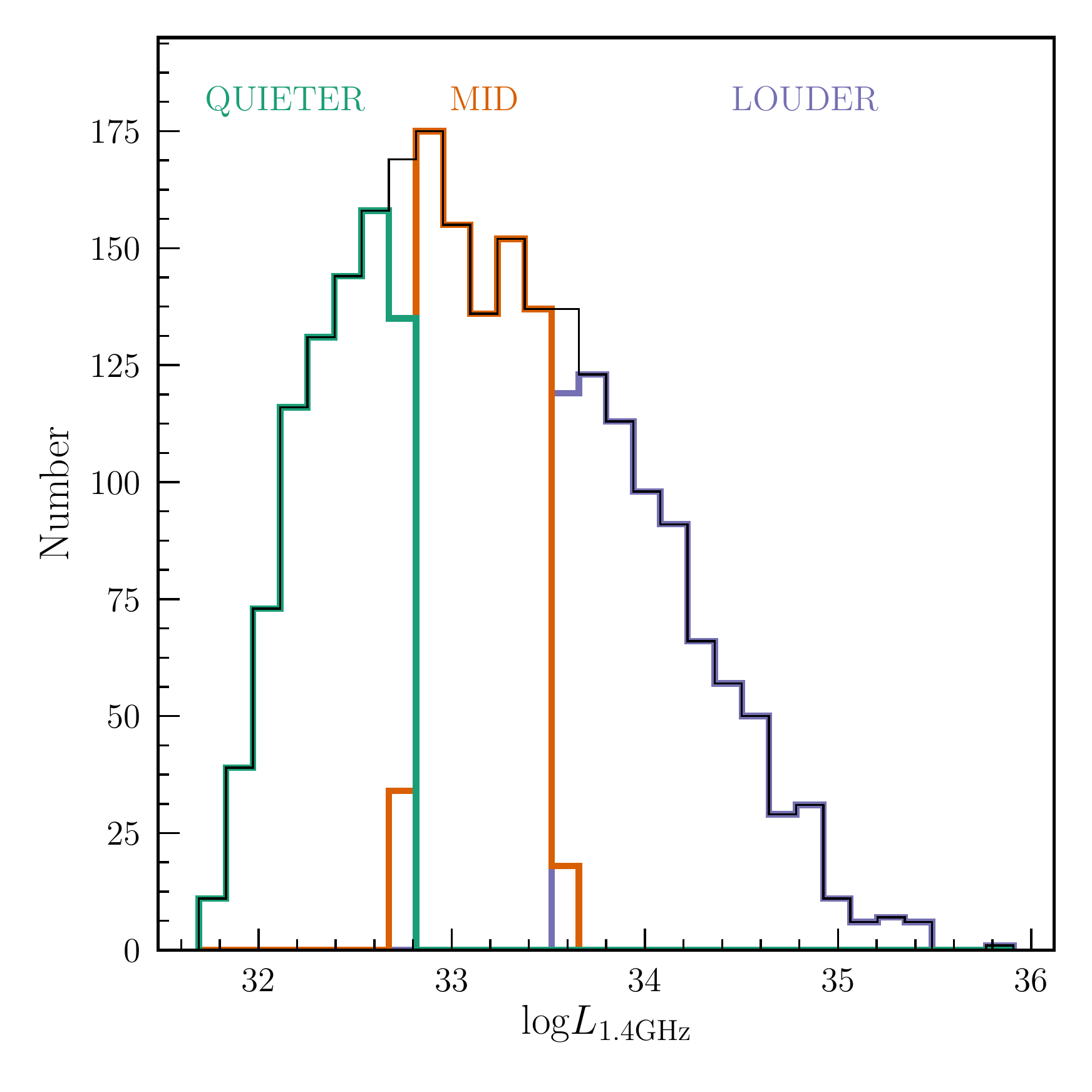}
	\end{subfigure}
	\hfill
	\begin{subfigure}{0.45\textwidth}
		\includegraphics[width=\textwidth]{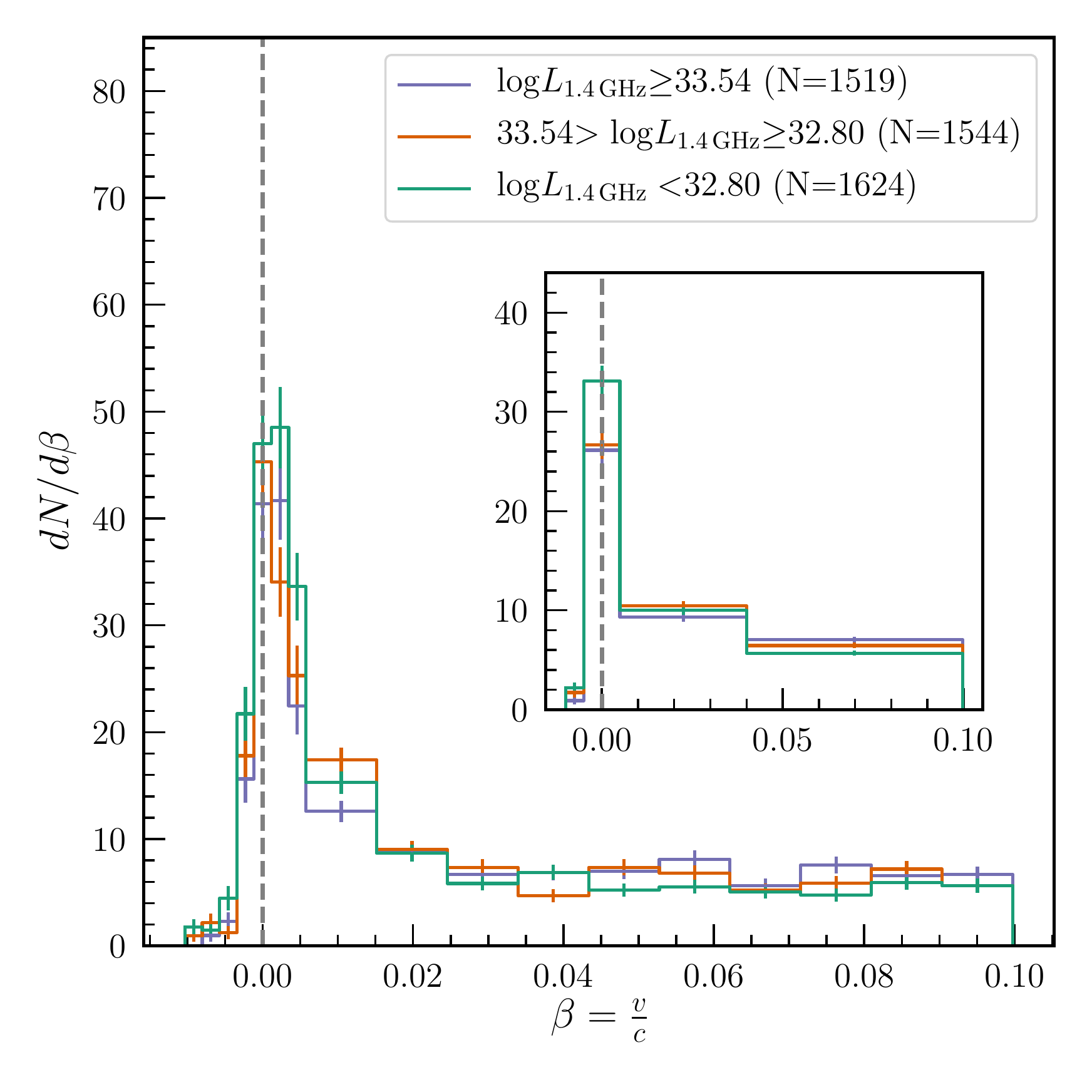}
	\end{subfigure}
	\caption{(\emph{Left}) Distribution of radio luminosity (1.4 GHz; FIRST).  The  solid-colored lines indicate a separation of the quasars into three equal groups.  (\emph{Right}) Absorber velocity distributions (radio luminosity groups).  We see no specific trend in the three groups, other than a small excess of intrinsic absorbers in radio quieter QSOs.}
	\label{fig:L_rad}
\end{figure*}

In addition to using the ratio of radio-to-optical brightness, ``radio loudness" can also be determined directly from an object's radio luminosity.  Upon separating the quasars instead into groups based on their FIRST 1.4\,GHz luminosity (Figure~\ref{fig:L_rad}, left panel) we find that there is a very slight excess of low velocity absorbers in less radio luminous quasars (Figure~\ref{fig:L_rad}, right panel)---opposite of our findings with $\alpha_{\rm ro}$.  At high velocity there do not appear to be any significant trends.  We will discuss the differences between our $\alpha_{\rm ro}$ and radio luminosity results further in Section~\ref{sec:disc_rad}.

\subsubsection{Radio Spectral Index} \label{sec:alpha_rad}

\begin{figure*}
	\centering
	\begin{subfigure}{0.45\textwidth}
		\includegraphics[width=\textwidth]{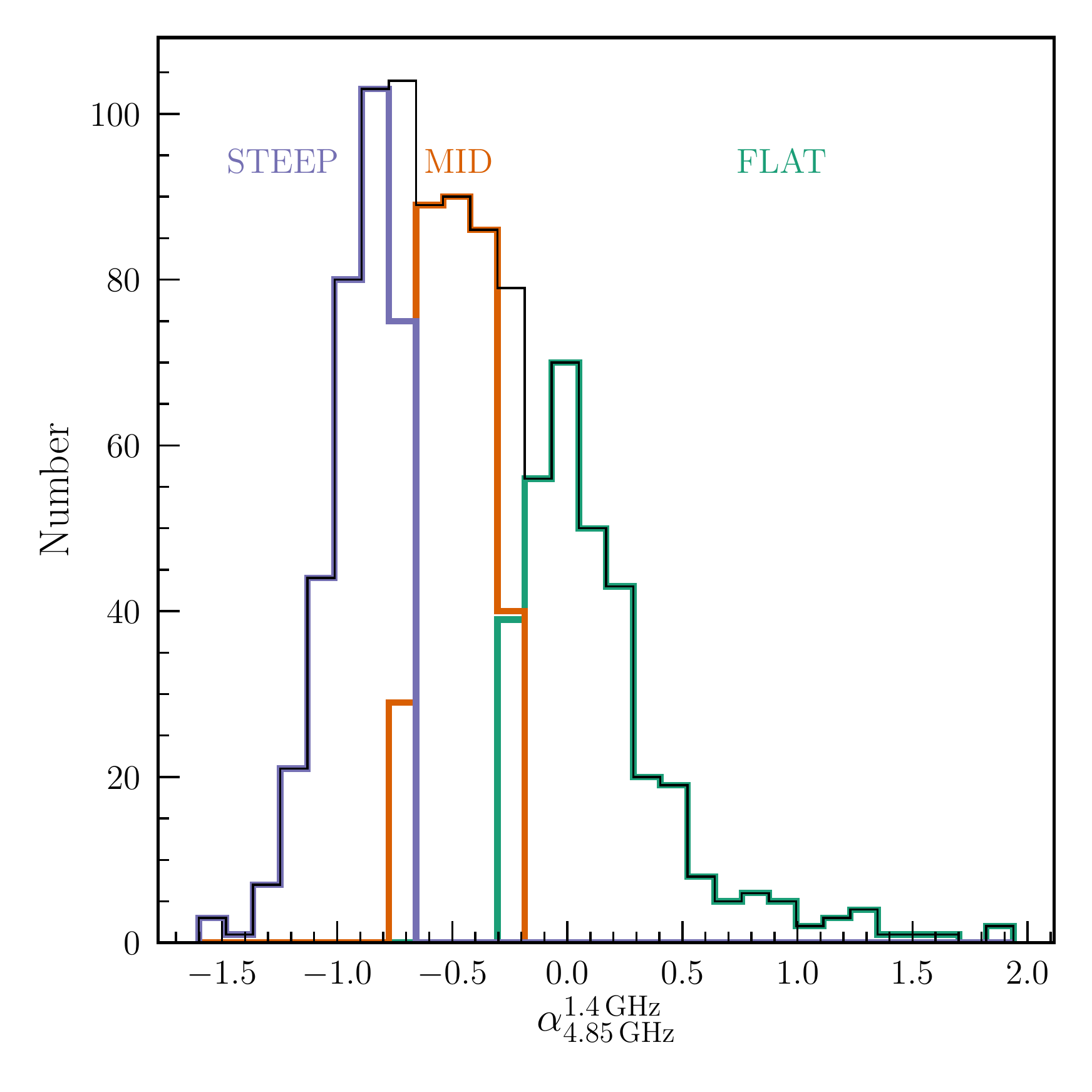}
	\end{subfigure}
	\hfill
	\begin{subfigure}{0.45\textwidth}
		\includegraphics[width=\textwidth]{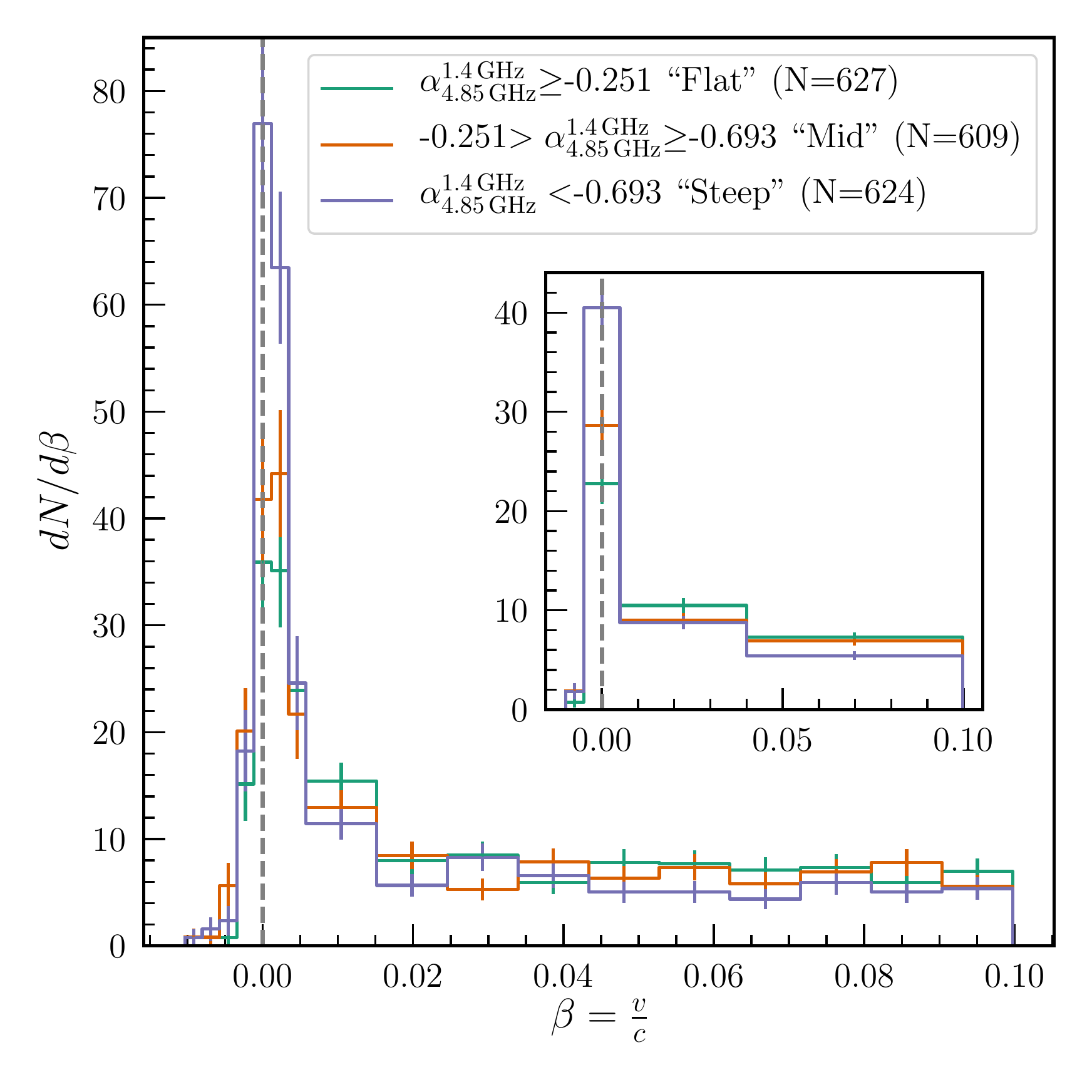}
	\end{subfigure}
	\caption{(\emph{Left}) Distribution of radio spectral indices ($\alpha^{1.4}_{4.85}$).  The solid-colored lines indicate a separation of the quasars into three equal groups. (\emph{Right}) Absorber velocity distributions (radio spectral index groups).  At low velocity there is an excess of intrinsic absorption in steep-spectrum objects with a clear downtrend towards flat-spectrum objects. At velocities exceeding $\sim\SI{1500}{\km\per\s}$ ($\beta\sim0.005$) flat-spectrum objects are (marginally) in excess of steep-spectrum in both the exponential tail and intervening components.}
	\label{fig:alpha20_6_3types}
\end{figure*}

In contrast to our inconclusive findings with respect to radio loudness, we find profound differences in the velocity distribution of NALs as a function of radio spectral index.  We split the quasars into three equal groups (left panel) and plotted the normalized velocity distributions (right panel); see Figure~\ref{fig:alpha20_6_3types}.  Consistent with previous work \citep[e.g.,][and references therein]{Richards2001b} we see a significant trend in the $v\sim0$ Gaussian peak with an increasing density of absorbers going from flat-spectrum to steep-spectrum quasars: the large excess of low-velocity absorbers in steep-spectrum quasars is highly significant. Conversely, there is an exponential tail of higher-velocity absorbers that seems to have a (marginal) excess in flat-spectrum quasars.  However, at the largest velocities, $v>\sim\SI{7500}{\km\per\s}$ we do not find a significant difference between the three groups, contrasting the results of \citet{Richards2001b}, where they found an excess of flat-spectrum, high-velocity absorbers.  The much larger data sample of our work has allowed us to use many more bins (thus higher-resolution) in the velocity distributions, compared to the two bins used in \citet{Richards2001b}.  Using the small number of bins seen in the inset, there is a slight, though not significant, excess in flat-spectrum high-velocity absorbers, which is not nearly as large a difference as in \citet{Richards2001b}.

\subsection{Insights from Optical Luminosity} \label{sec:M_i}

\begin{figure*}
	\centering
	\begin{subfigure}{0.45\textwidth}
		\includegraphics[width=\textwidth]{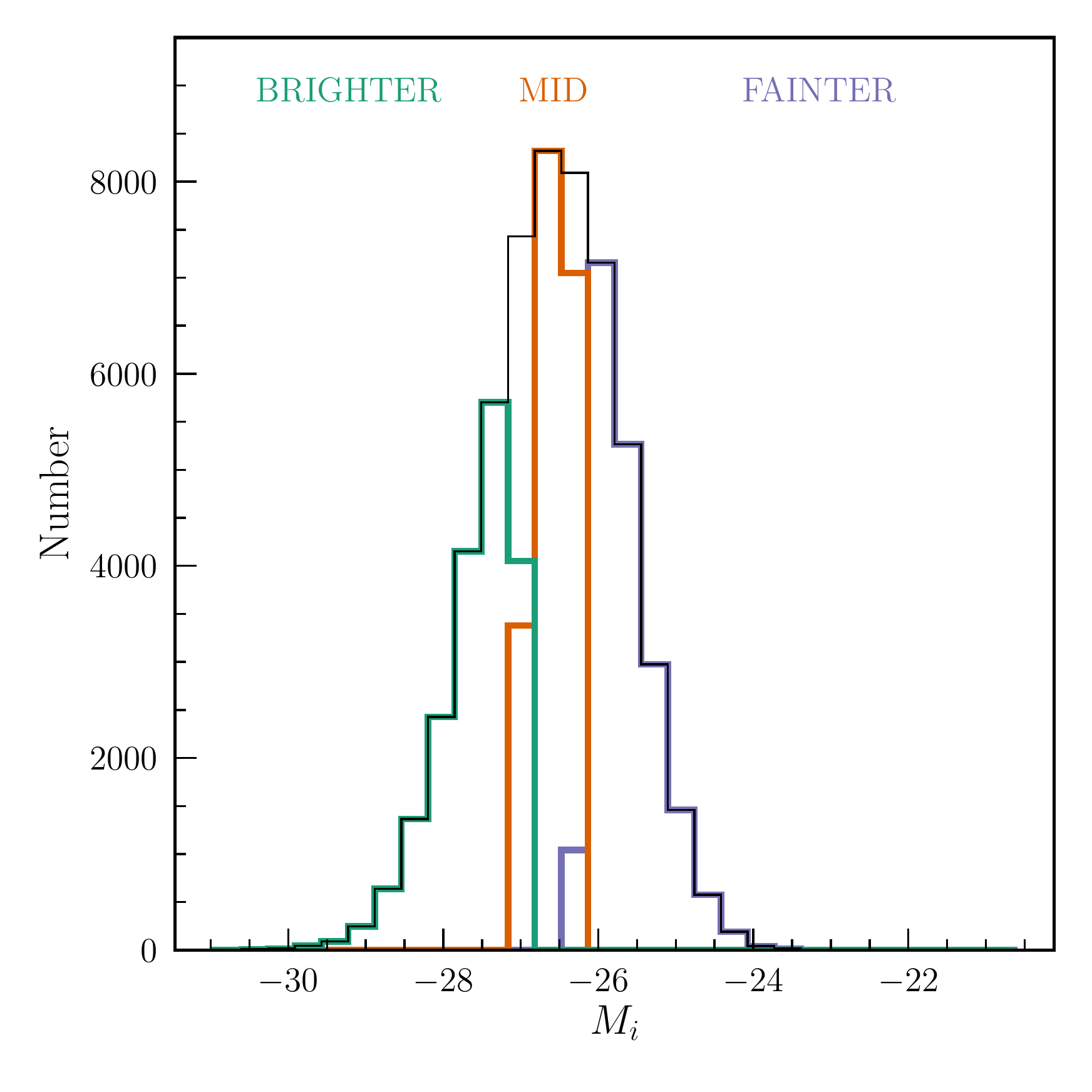}
	\end{subfigure}
	\hfill
	\begin{subfigure}{0.45\textwidth}
		\includegraphics[width=\textwidth]{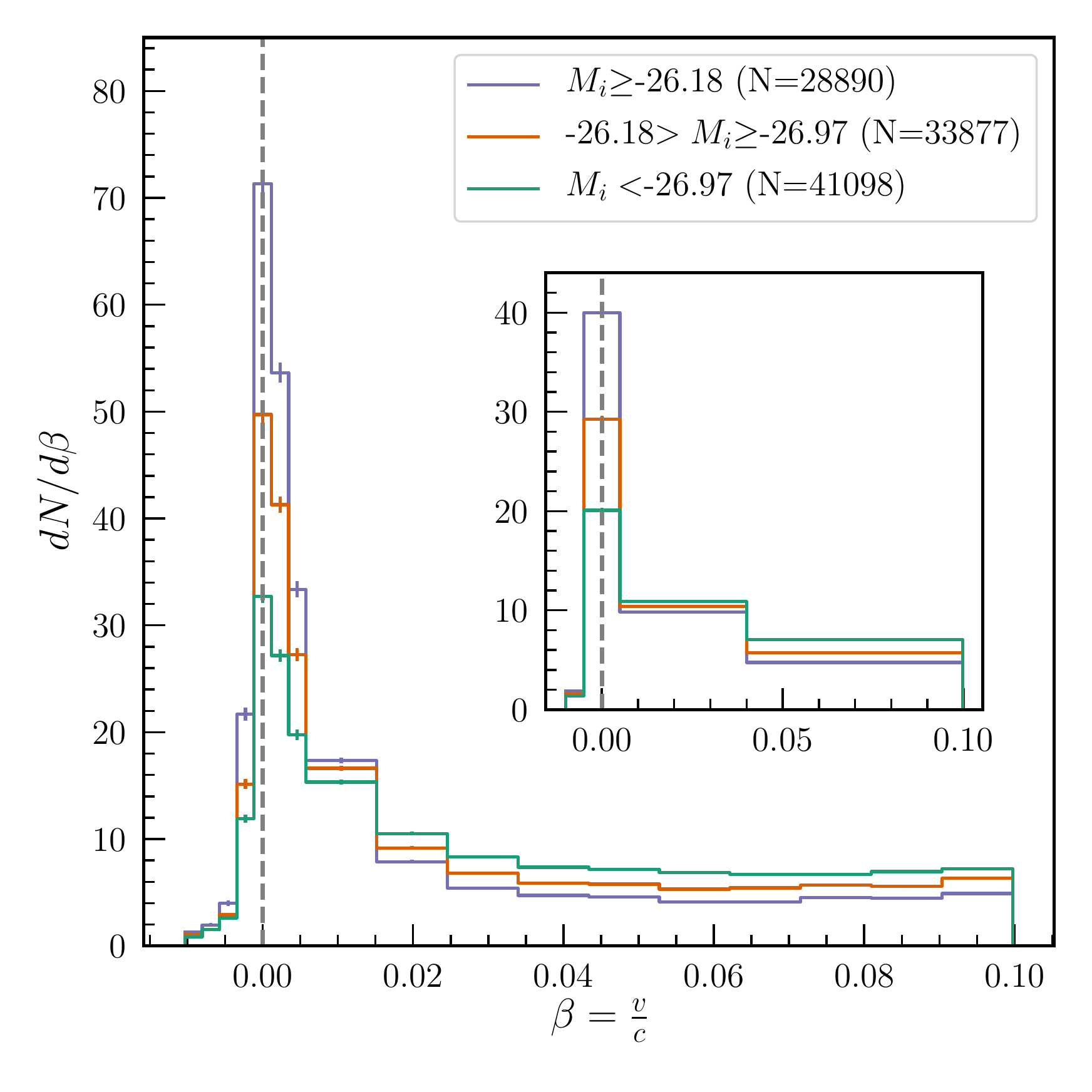}
	\end{subfigure}
	\caption{(\emph{Left}) Distribution of absolute magnitudes ($M_i$).  The solid-colored lines indicate a separation of the quasars into three equal groups.  (\emph{Right}) Absorber velocity distributions (absolute magnitude groups).  The green histogram corresponds to low luminosity objects, while the purple objects are high luminosity objects.  A significant anti-correlation of increasing associated absorption with decreasing luminosity is observed at low velocity, while the opposite is true at higher velocities.  This result would seemingly agree with the concept of a radiation line-driven wind:  more luminous objects are capable of driving high velocity absorbers, while fainter objects are not and cause absorbers to pile up at low velocities.}
	\label{fig:abs_mag}
\end{figure*}

It is important to consider whether the strong differences in the low-velocity absorbers as a function of radio spectral index are truly trends in radio spectral index and not something else.  Since beaming of flat-spectrum sources could make them brighter in the optical (and thus more likely to be selected) and higher S/N in brighter objects could make it easier to identify weak absorption features, we next consider the NAL distribution as a function of optical luminosity, here as measured by the absolute magnitude, $M_i$.  When the absorber sample is split by absolute magnitude we see a significant correlation between their velocities and optical luminosity: lower luminosity quasars are more likely to have associated absorption, but higher luminosity quasars have an excess of high-velocity absorbers (Figure~\ref{fig:abs_mag}).  The differences seen between high- and low-luminosity quasars is similar to that seen between flat- and steep-spectrum quasars.

\begin{figure*}
	\centering
	\includegraphics[width=0.55\textwidth]{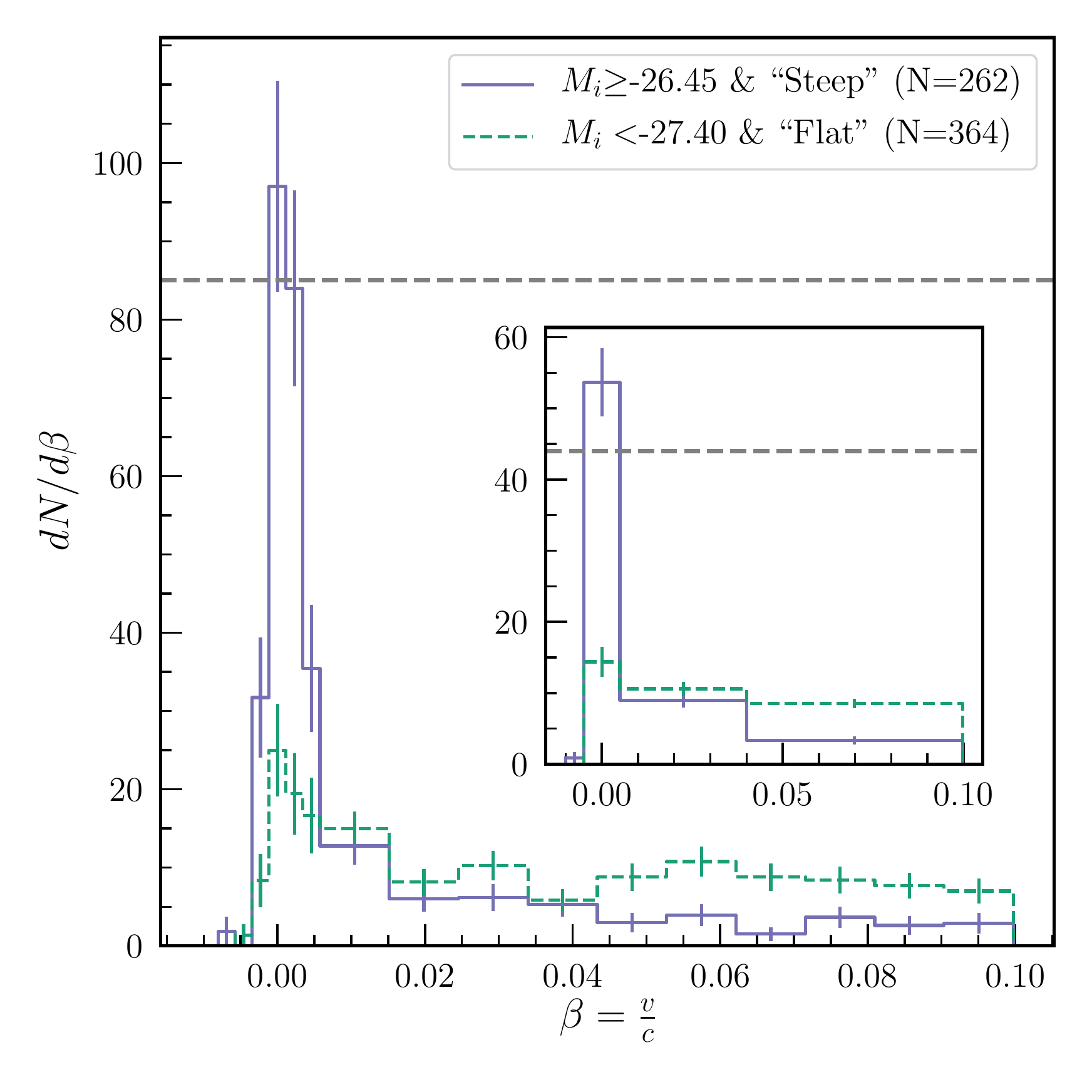}
	\caption{Absorber velocity distributions ($\alpha_{\rm rad}$ and $M_i$ groups).  This plot combines $\alpha_{\rm rad}$ and $M_i$ into six different absorber groups (as described in the text), though for clarity just the two extremes of these six groups are included, and a clear trend emerges.  Optically dim, steep-spectrum quasars have an excess of low-velocity, and a lack of high-velocity, absorption, while the exact opposite is true for luminous, flat-spectrum quasars.  This result could be seen to support the notion of failed (and/or polar) winds in low-luminosity objects which do not have the necessary radiation power to drive strong equatorial winds as their brighter siblings.  More fundamentally, this result supports the hypothesis that both driving power (i.e., luminosity), and orientation (i.e., radio spectral index) contribute to the velocity distribution of NALs.}
	\label{fig:alpha20_6_Mi}
\end{figure*}

Such a scenario still leaves the question of whether the strong trends seen in the velocity distributions as functions of optical luminosity and radio spectral index are correlated, potentially from a correlation of those intrinsic properties.  As such, we took the three $M_i$ groups, as above, and split each in half by $\alpha_{\rm rad}$, creating a ``steep'' and ``flat'' group for each absolute magnitude group.  Examining the velocity distributions of these six groups, an equally strong trend is observed, thus the optical luminosity trend is unlikely to be driving the spectral-index trend.  Indeed, the most significant results are found by combining these properties.  In Figure~\ref{fig:alpha20_6_Mi} we see that low-luminosity, steep-spectrum quasars have such a large excess of AALs that we had to change the plot limit from 75 to 105, whereas there is virtually no excess of AALs in high-luminosity, flat-spectrum quasars.  On the other hand, at high velocity there is a significant excess of NALs in high-L, flat-spectrum sources as compared to low-L, steep-spectrum sources.  \citet{Richards2001b} argued that such differences at high velocity can be used to statistically determine the fraction of intrinsic NALs that masquerade as intervening systems.  Assuming that the mean level of high-velocity NALs in low-L, steep-spectrum sources are entirely intervening systems, then the fraction of intrinsic \ion{C}{IV} NALs can be estimated from the excess seen in high-L, flat-spectrum.  The result of such an analysis with our sample yields an estimate of \SI[separate-uncertainty = true, multi-part-units = repeat]{30.4\pm6.4}{\percent} as compared to \SI{36}{\percent} found by \citet{Richards2001b}.

Since there is a strong optical/UV luminosity dependence to the NAL distribution, we investigate what effect marginalizing over optical/UV luminosity has on our radio results ($\alpha_{\rm rad}$, $\alpha_{\rm ro}$, and $L_{\rm 1.4\,GHz}$).  To do this we simply gridded the luminosity-radio space and calculated the absolute magnitude distribution of each cell, found the group which had the smallest number of objects in each bin, and then randomly sampled the other two groups, restricting them to have only as many objects as that smallest number.  We randomly sampled the groups that had more than the minimum in each bin, repeating this process 1000 times and took the average result.  

\begin{figure*}
    \centering
    \begin{subfigure}{0.32\textwidth}
		\includegraphics[width=\textwidth, keepaspectratio]{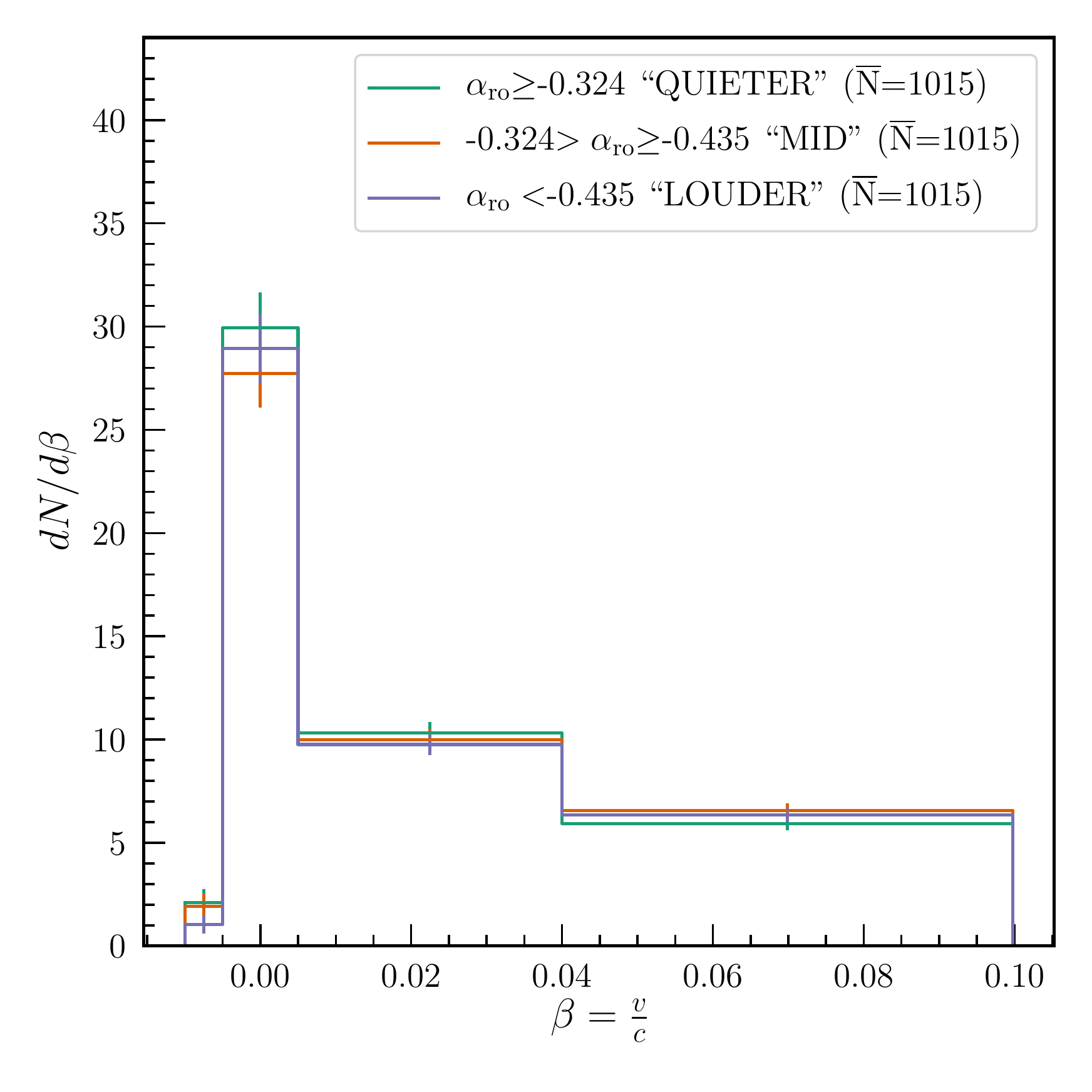}
	\end{subfigure}
	\hfill
	\begin{subfigure}{0.32\textwidth}
		\includegraphics[width=\textwidth, keepaspectratio]{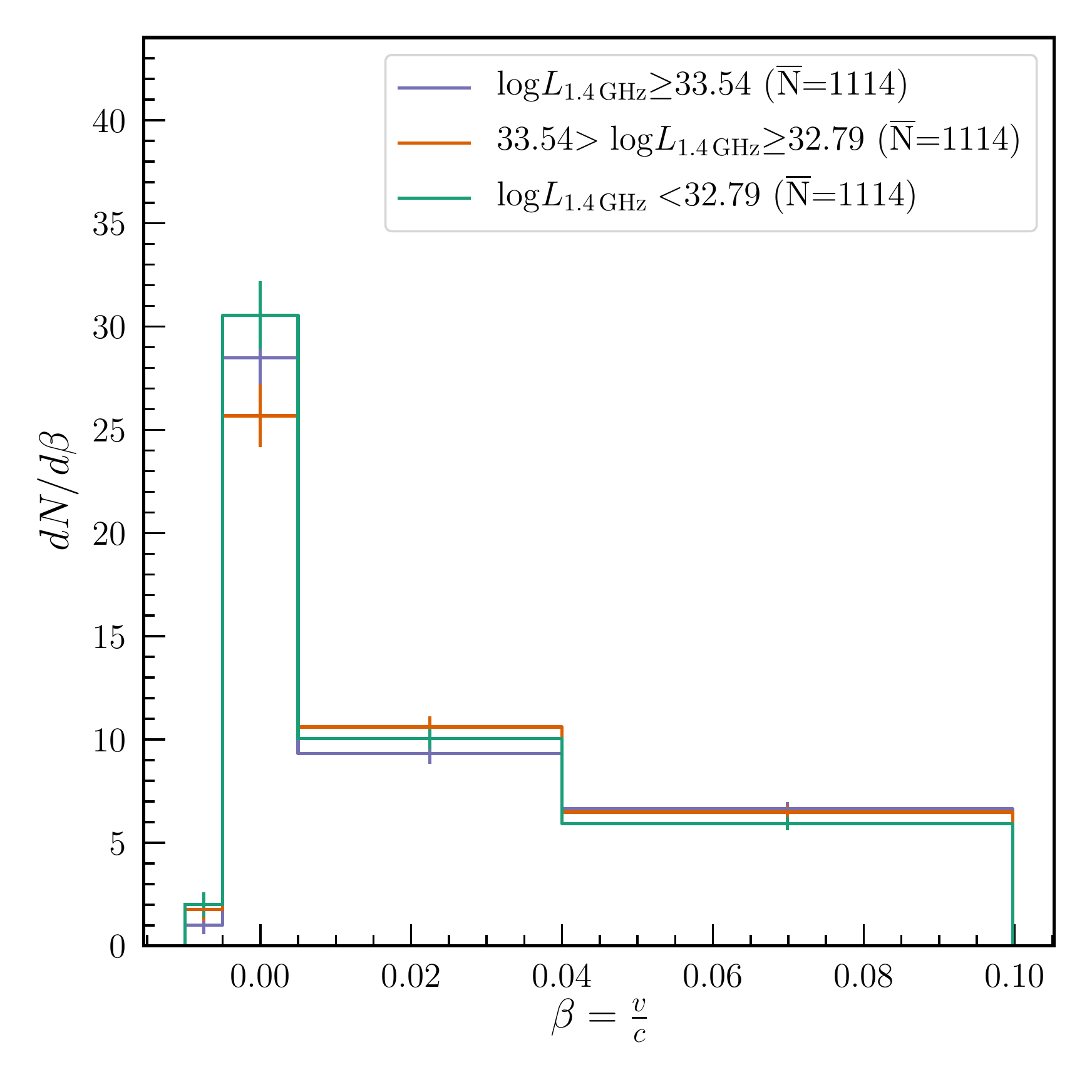}
	\end{subfigure}
	\hfill
	\begin{subfigure}{0.32\textwidth}
		\includegraphics[width=\textwidth, keepaspectratio]{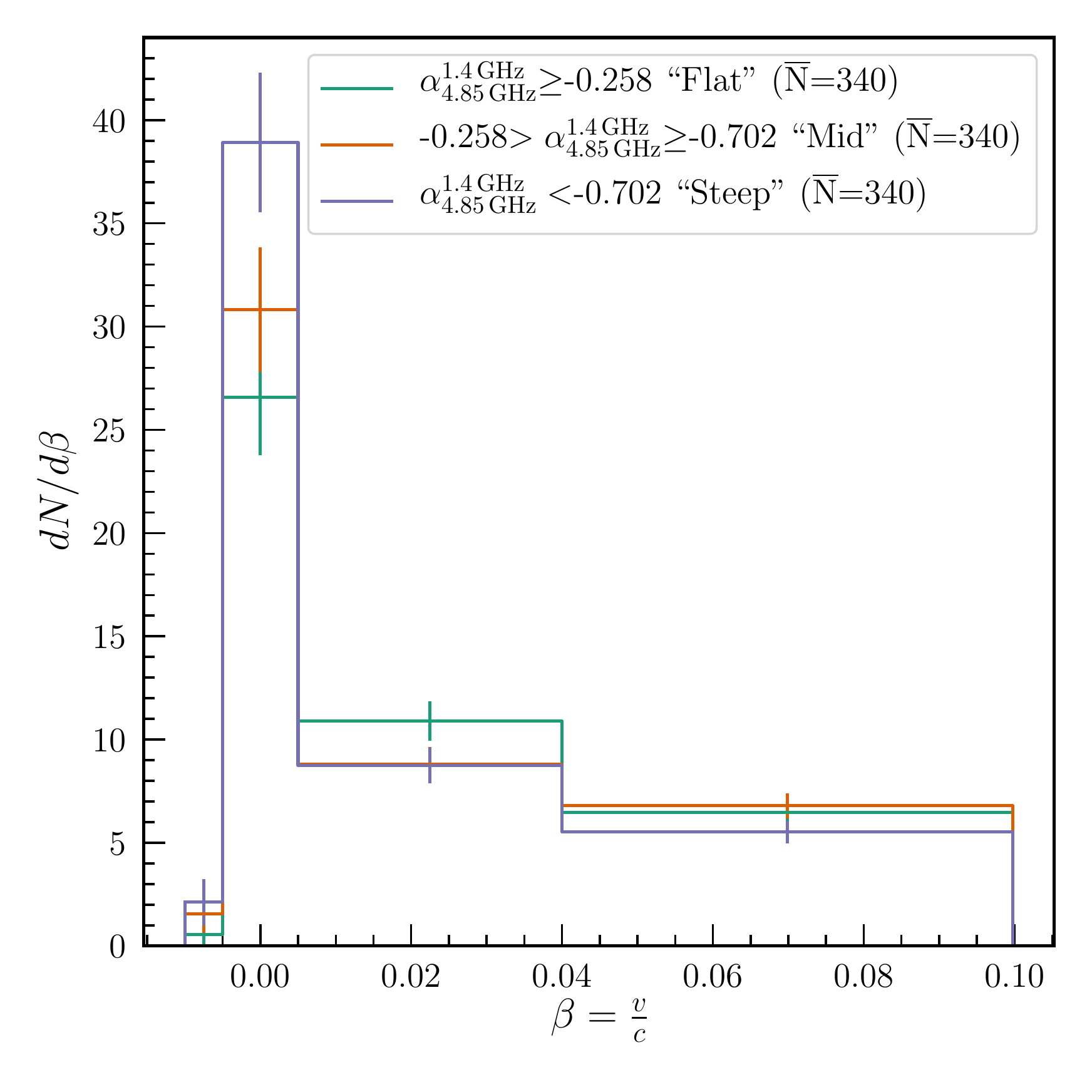}
	\end{subfigure}
    \caption{Absorber velocity distributions with matched $M_i$ as functions of $\alpha_{\rm ro}$ (\emph{Left}), $L_{\rm 1.4\,GHz}$ (\emph{Middle}), and $\alpha_{\rm rad}$ (\emph{Right}).  The weak trends observed in Figures~\ref{fig:alpha_ro_wNonRad} and \ref{fig:L_rad} are no longer seen in the left and middle plots respectively, suggesting that the apparent differences in the NAL distribution as a function of radio loudness may instead be an effect of optical (rather than radio) luminosity.  While marginalizing the radio spectral index groups over optical/UV luminosity has weakened the trends observed in Figure~\ref{fig:alpha20_6_3types}, those trends are still visible and relatively significant.  This weakening suggests that the $\alpha_{\rm rad}$ trends are partially caused by optical luminosity, but not entirely, and therefore that the $\alpha_{\rm rad}$ trends have a unique physical origin.}
    \label{fig:beta_comp_Mi}
\end{figure*}

Figure~\ref{fig:beta_comp_Mi} plots the average absorber velocity distributions of the 1000 runs for the normalized groups of $\alpha_{\rm ro}$, $L_{\rm 1.4\,GHz}$, and $\alpha_{\rm rad}$, (left, middle, and right panels, respectively).  For both $\alpha_{\rm ro}$ and radio luminosity any trends we observed previously are now no longer seen, suggesting that any differences seen in the NAL distribution as a function of those radio properties were instead due to differences in the optical/UV luminosity distribution.  Furthermore, for both properties the so-called ``mid" group now shows the least absorption in the AAL bin, changing the perceived trend (or lack thereof).  These results are discussed further in the context of the origin of the radio emission in Section~\ref{sec:disc_rad}.  However, the correlations observed in the absorber velocity distributions for radio spectral index were largely unaffected by the normalized optical luminosity (the trends are somewhat weakened but are still visible and relatively significant\footnote{The much larger errorbars in the $\alpha_{\rm rad}$ plot are due to the decreased number of absorbers in the plot.  Radio spectral indices from CNSS and VLASS will significantly improve the errorbars for this kind of analysis.}), indicating that the distribution of NALs as a function of radio spectral index of quasars is not dominated by the optical luminosity, and that the strong correlations we see between intrinsic absorption and radio spectral index are real.  These results are summarized in Table~\ref{tab:comp_Mi}, which tabulates the difference (and significance thereof) of ${\rm dN}/{\rm d}\beta$ between the class extrema of the three data types shown in Figure~\ref{fig:beta_comp_Mi}, both before and after marginalizing over optical/UV luminosity.

\begin{table*}
    \caption{${\rm d}N/{\rm d}\beta$ Differences Between Radio Property Extrema}
    \label{tab:comp_Mi}
    \begin{tabular}{ccccccccccccc}
    \hline
    & \multicolumn{6}{c}{Before Marginalization} & \multicolumn{6}{c}{After Marginalization} \\
    \cline{2-13}
    Intrinsic Property & \multicolumn{2}{c}{Gaussian} & \multicolumn{2}{c}{Exponential} & \multicolumn{2}{c}{Intervening} & \multicolumn{2}{c}{Gaussian} & \multicolumn{2}{c}{Exponential} & \multicolumn{2}{c}{Intervening} \\
    \hline
    $\alpha_{\rm ro}$ & 4.52 & 2.21$\sigma$ & 0.87 & 1.36$\sigma$ & 0.21 & 0.54$\sigma$ & 1.02 & 0.42$\sigma$ & 0.57 & 0.75$\sigma$ & 0.43 & 0.94$\sigma$ \\
    $L_{\rm 1.4\,GHz}$ & 6.99 & 3.40$\sigma$ & 0.69 & 1.09$\sigma$ & 1.36 & 3.38$\sigma$ & 2.06 & 0.89$\sigma$ & 0.74 & 1.05$\sigma$ & 0.72 & 1.61$\sigma$ \\
    $\alpha_{\rm rad}$ & 17.76 & 5.22$\sigma$ & 1.74 & 1.71$\sigma$ & 1.87 & 2.87$\sigma$ & 12.35 & 2.81$\sigma$ & 2.15 & 1.66$\sigma$ & 0.95 & 1.18$\sigma$ \\
    $M_i$ & 19.91 & 43.59$\sigma$ & 1.04 & 7.44$\sigma$ & 2.28 & 27.75$\sigma$ & N/A & N/A & N/A & N/A & N/A & N/A \\
    \hline
    \end{tabular}
\end{table*}

The importance of marginalizing over luminosity cannot be emphasized enough as many of the discrepant claims in the literature may well have an origin in sample differences.  For example, \citet{Vestergaard2003} argues that sample differences may explain contradictory results with \citet{Baker2002} and possibly \citet{Ganguly2001}.

\subsection{Comparison to \citet{Richards2001b}} \label{sec:gtr_comp}

Having now created our own plots of ${\rm dN}/{\rm d}\beta$ as functions of both radio loudness and radio spectral index, we seek to test the findings of \citet{Richards2001b} (the former is seen in their Figures~28 \& 29, the latter in their Figures~31 \& 32), which is part of the driving motivation for this paper.  Concerning radio loudness, they found a small excess of radio quiet objects at high velocity, and no trend at low velocity. This contrasts with our initial findings (before marginalizing over luminosity) in Section~\ref{sec:radio_loudness} where we saw a slight excess (albeit at very low significance) of AALs in radio louder objects, and no trend at high velocity.    Our results for $\alpha_{\rm rad}$, seen in Section~\ref{sec:alpha_rad}, are in far better agreement with \citet{Richards2001b}, showing an excess of low-velocity absorbers in steep-spectrum sources and an excess of high-velocity absorbers in flat-spectrum sources.  It is important to note that conclusions drawn about the comparison of these results could be affected by differences in the samples used.  In fact, in sections~\ref{sec:radio_loudness}\&\ref{sec:alpha_rad} we discussed how the observed radio loudness trends were probably an artifact of optical luminosity trends (radio spectral showed no such dependence).  In this section we detail the methods we used to compare our results with \citet{Richards2001b} more accurately and discuss them in the context of the full paper.  

\begin{figure*}
    \centering
	\begin{subfigure}{0.35\textwidth}
		\includegraphics[width=\textwidth, keepaspectratio]{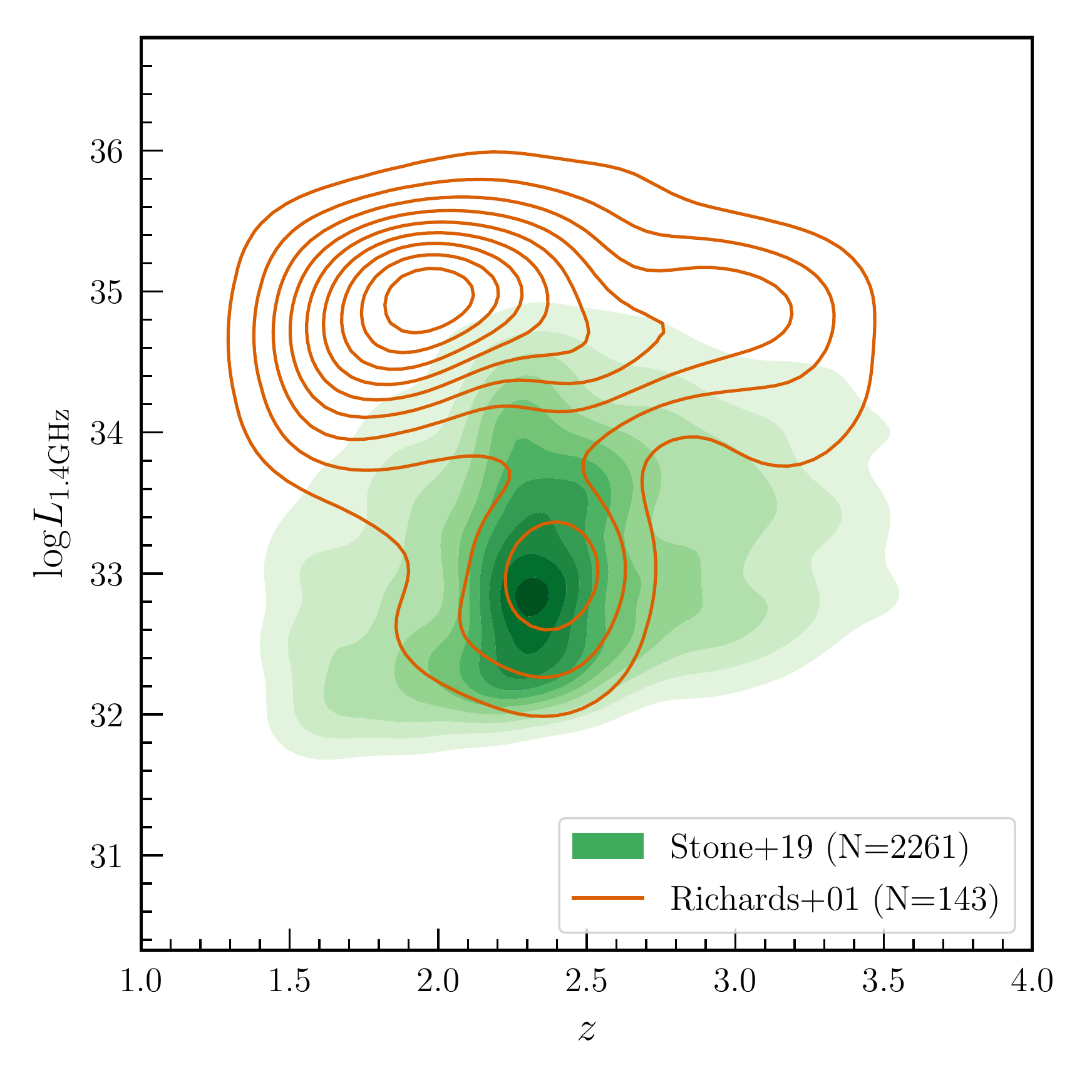}
	\end{subfigure}
	\quad
	\begin{subfigure}{0.35\textwidth}
		\includegraphics[width=\textwidth, keepaspectratio]{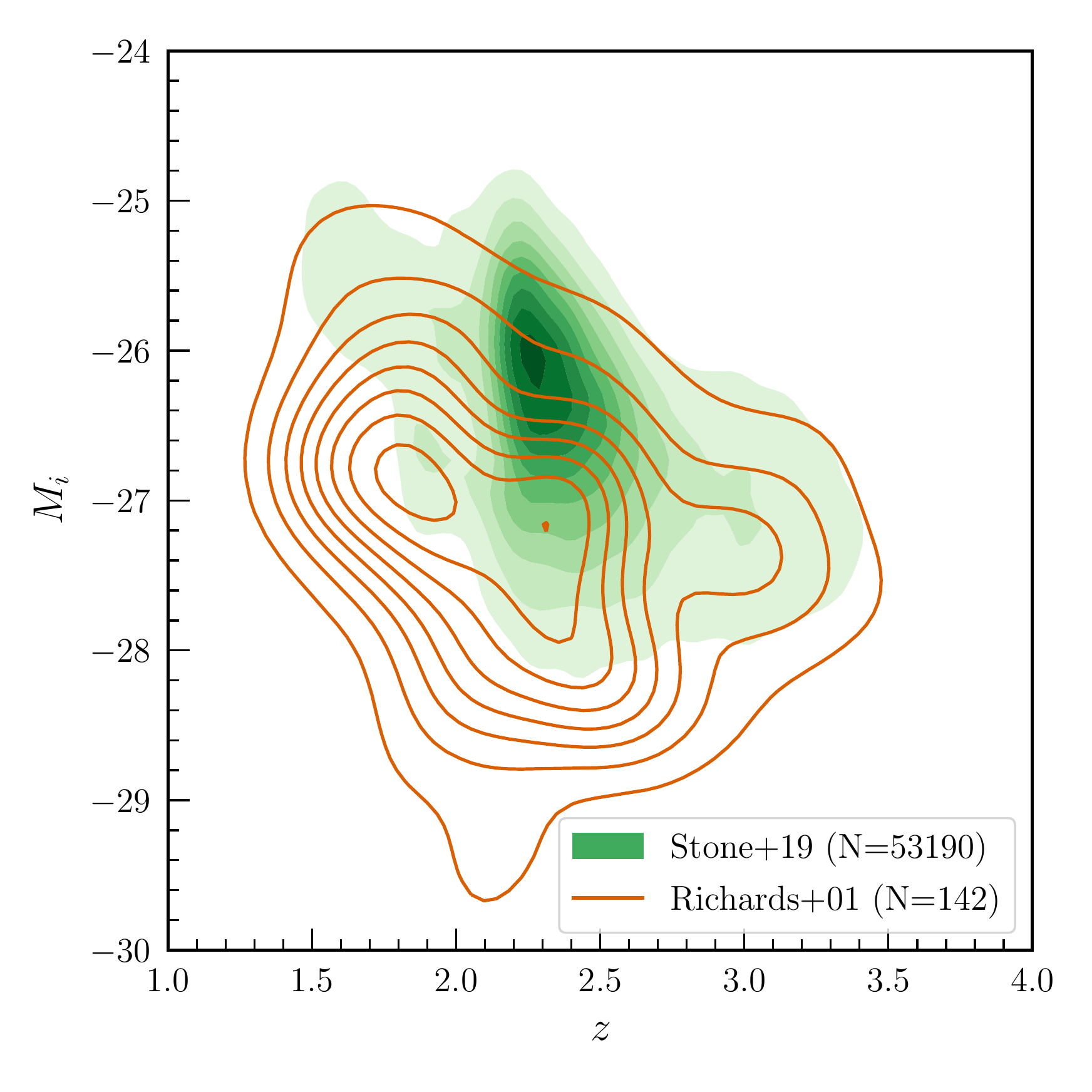}
	\end{subfigure}
	\\
	\begin{subfigure}{0.35\textwidth}
		\includegraphics[width=\textwidth, keepaspectratio]{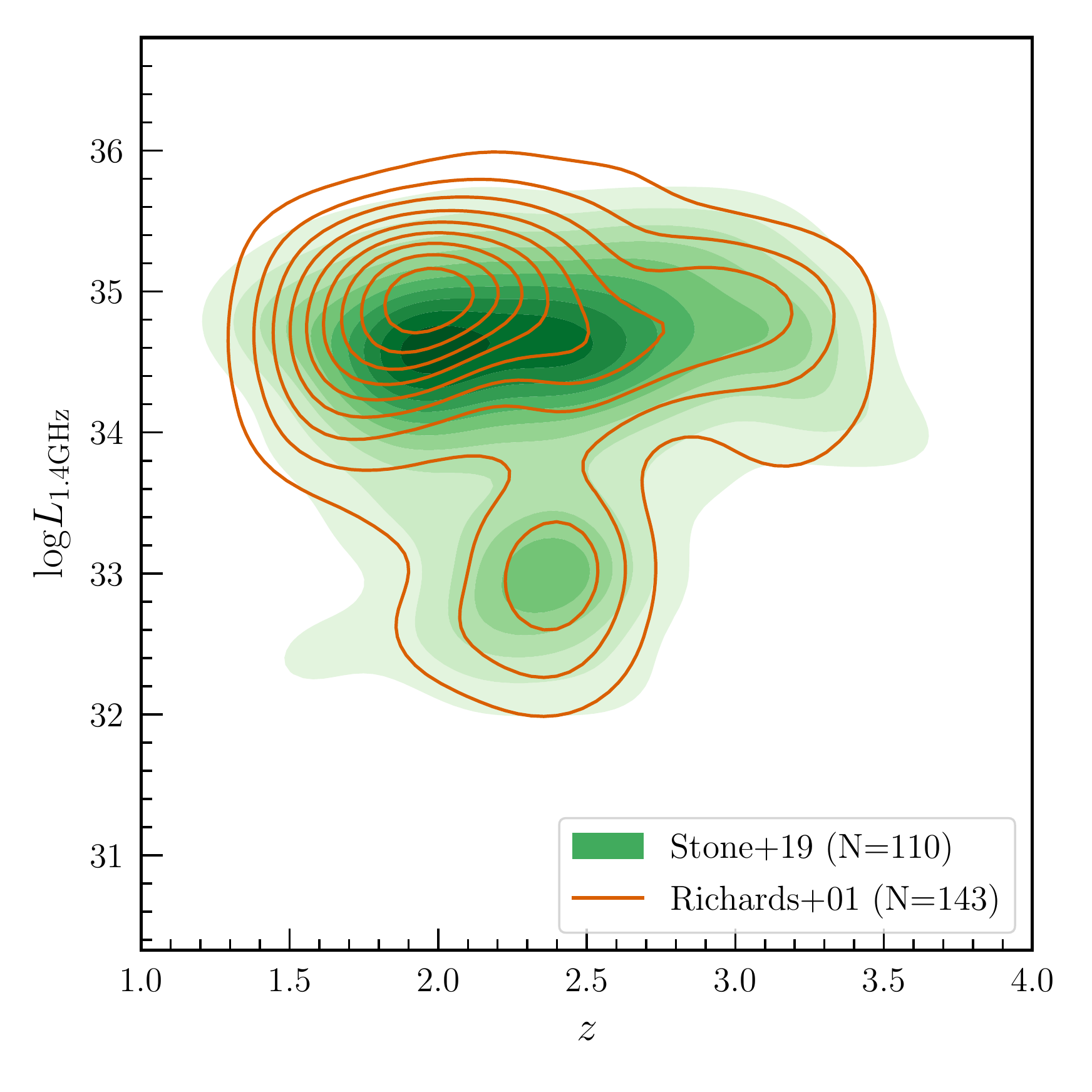}
	\end{subfigure}
	\quad
	\begin{subfigure}{0.35\textwidth}
		\includegraphics[width=\textwidth, keepaspectratio]{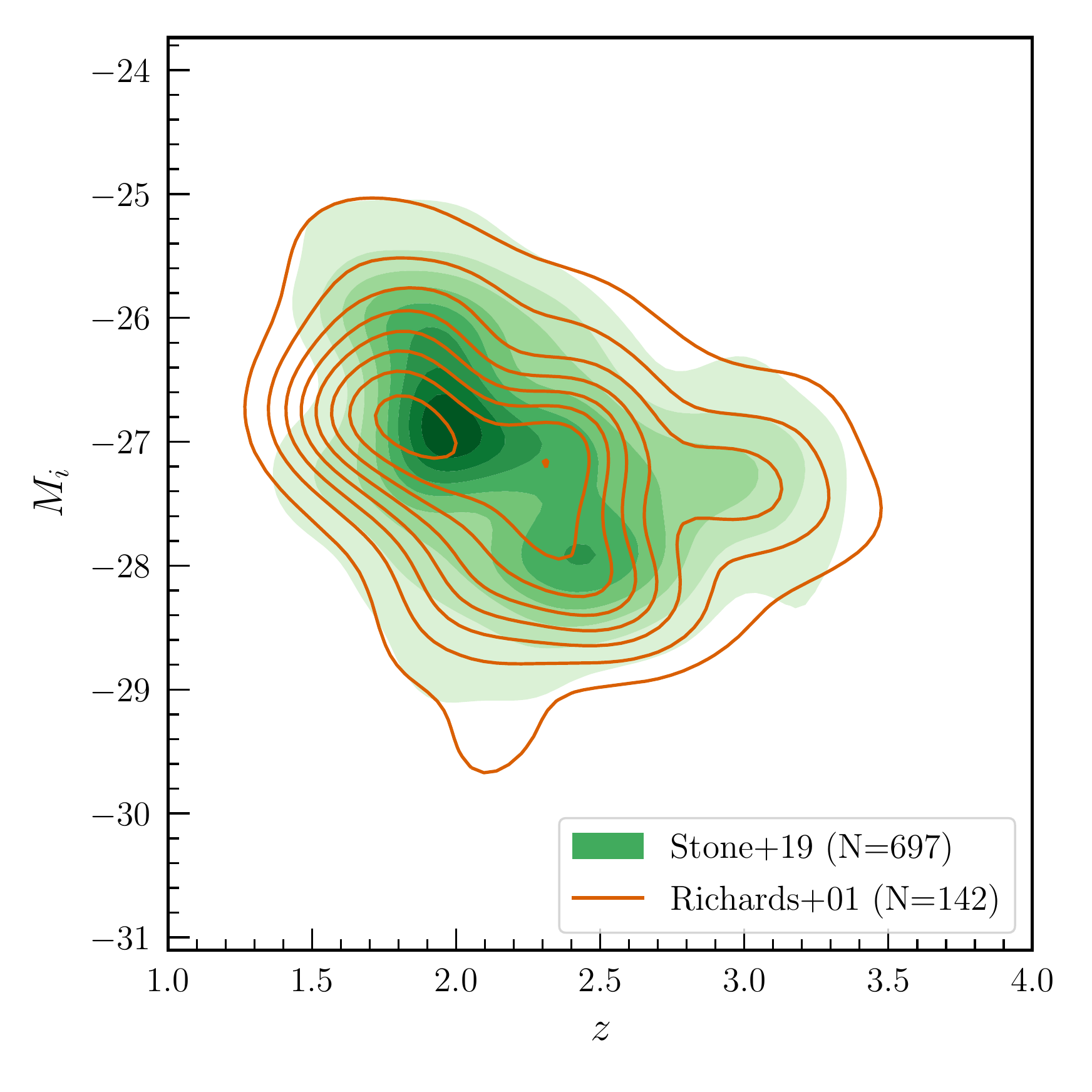}
	\end{subfigure}
    \caption{(\emph{Left}) $L_{\rm 1.4\,GHz}$ vs. $z$.  (\emph{Right}) $M_i$ vs. $z$.  Kernel density estimations of radio luminosity and optical magnitude vs redshift for our sample and the sample from \citet{Richards2001b}.  (\emph{Top Row}) We see significant differences in the distributions of these properties between the two samples.  (\emph{Bottom Row}) The matched sub-samples of our quasars better fit the quasars of \citet{Richards2001b} in the 2-D parameter spaces.}
	\label{fig:kde_gtr}
\end{figure*}

It would be dangerous to conclude that a feature seen by \citet{Richards2001b} (or one seen by this work) is not real if the samples were significantly different. The SDSS datasets presented herein contain information on thousands more quasars (and absorbers) than analyzed in \citet{Richards2001b}.  To more precisely compare the samples, it is essential to determine the extent to which the two samples probed the same quasar parameter spaces.  First, each sample was binned over redshift and radio luminosity ($L_{\mathrm{1.4GHz}}$), and then secondly, and separately, over redshift and absolute magnitude ($M_i$).  In Figure~\ref{fig:kde_gtr}, kernel density estimates of these two 2-D parameter spaces reveal significant differences between our current sample of quasars and that used by \citet{Richards2001b}.

\begin{figure*}
	\centering
	\begin{subfigure}{0.45\textwidth}
		\includegraphics[width=\textwidth]{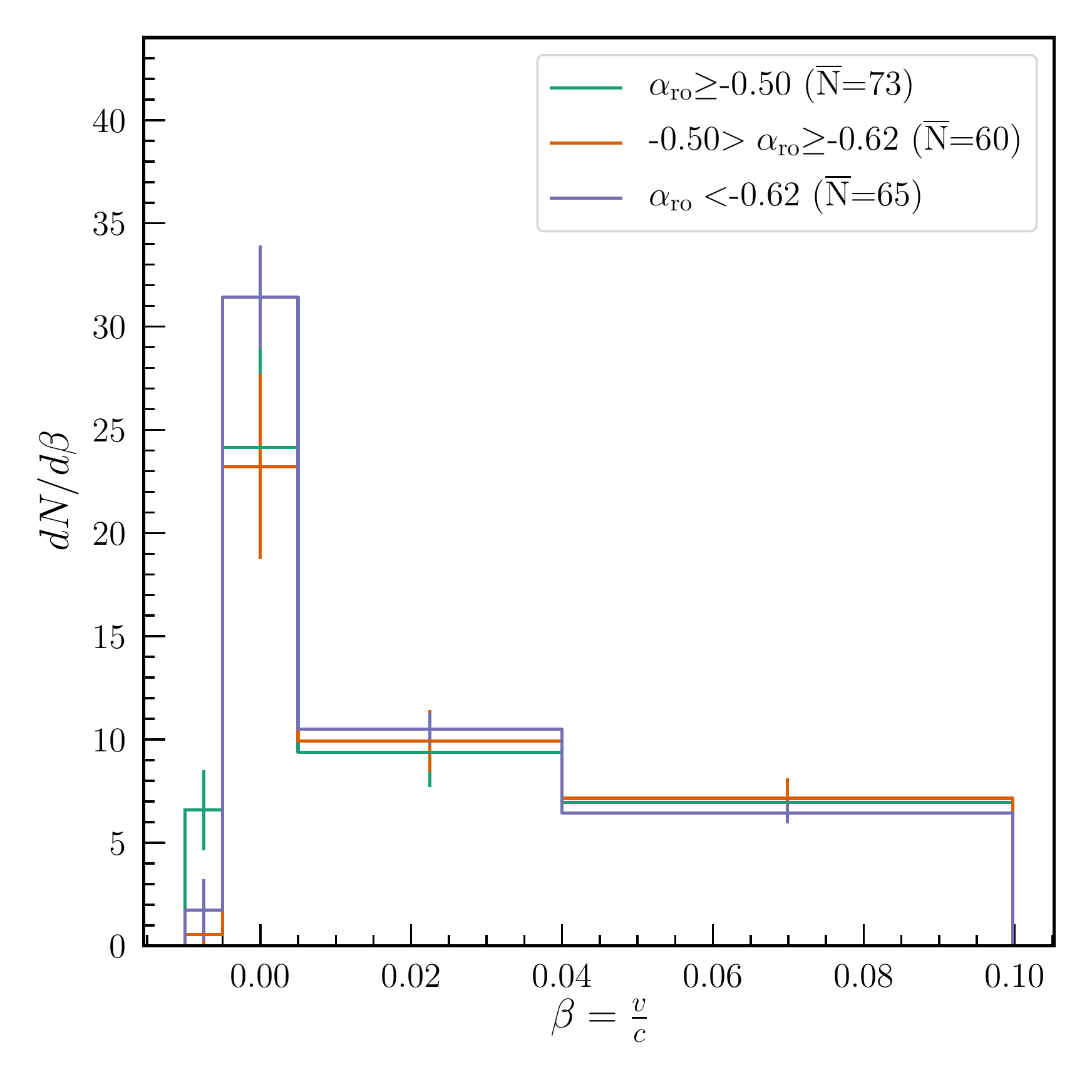}
	\end{subfigure}
	\hfill
	\begin{subfigure}{0.45\textwidth}
		\includegraphics[width=\textwidth]{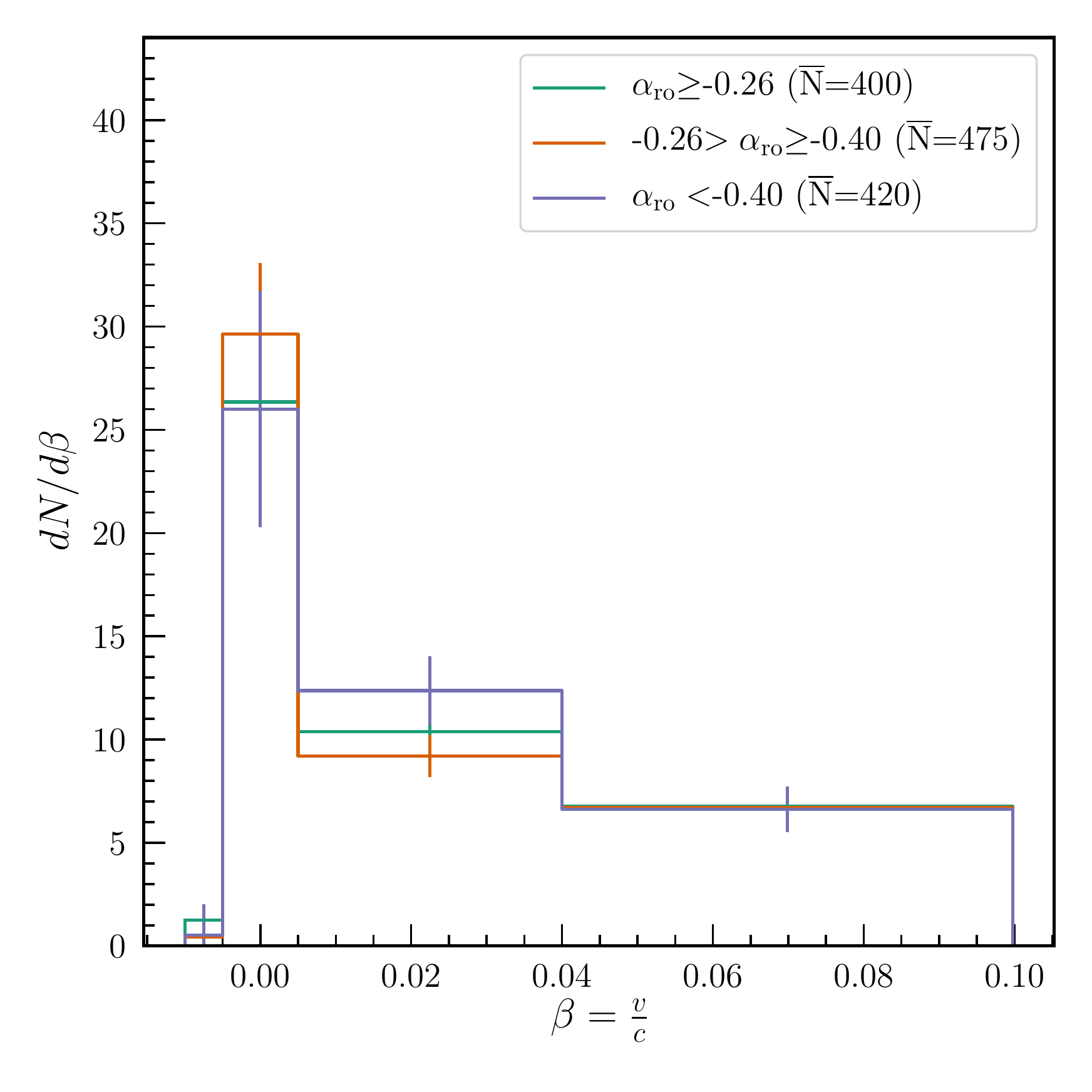}
	\end{subfigure}
    \caption{Average absorber velocity distributions as a function of RO spectral index ($\alpha_{\rm ro}$) for the matched sub-samples (\emph{left}, $L_{\mathrm{1.4GHz}}$-$z$; \emph{right}, $M_i$-$z$). The left plot shows a slight similarity to our Figure~\ref{fig:alpha_ro_wNonRad}.  In the plot on the right, conversely, there appears to be no trend in intrinsic absorption as a function $\alpha_{\rm ro}$.  Both of these plots show trends (or lack of trends) that are considerably different from Figures~28 \& 29 of \citet{Richards2001b}}
    \label{fig:beta_gtr_alpha_ro}
\end{figure*}

\begin{figure*}
	\centering
	\begin{subfigure}{0.45\textwidth}
		\includegraphics[width=\textwidth]{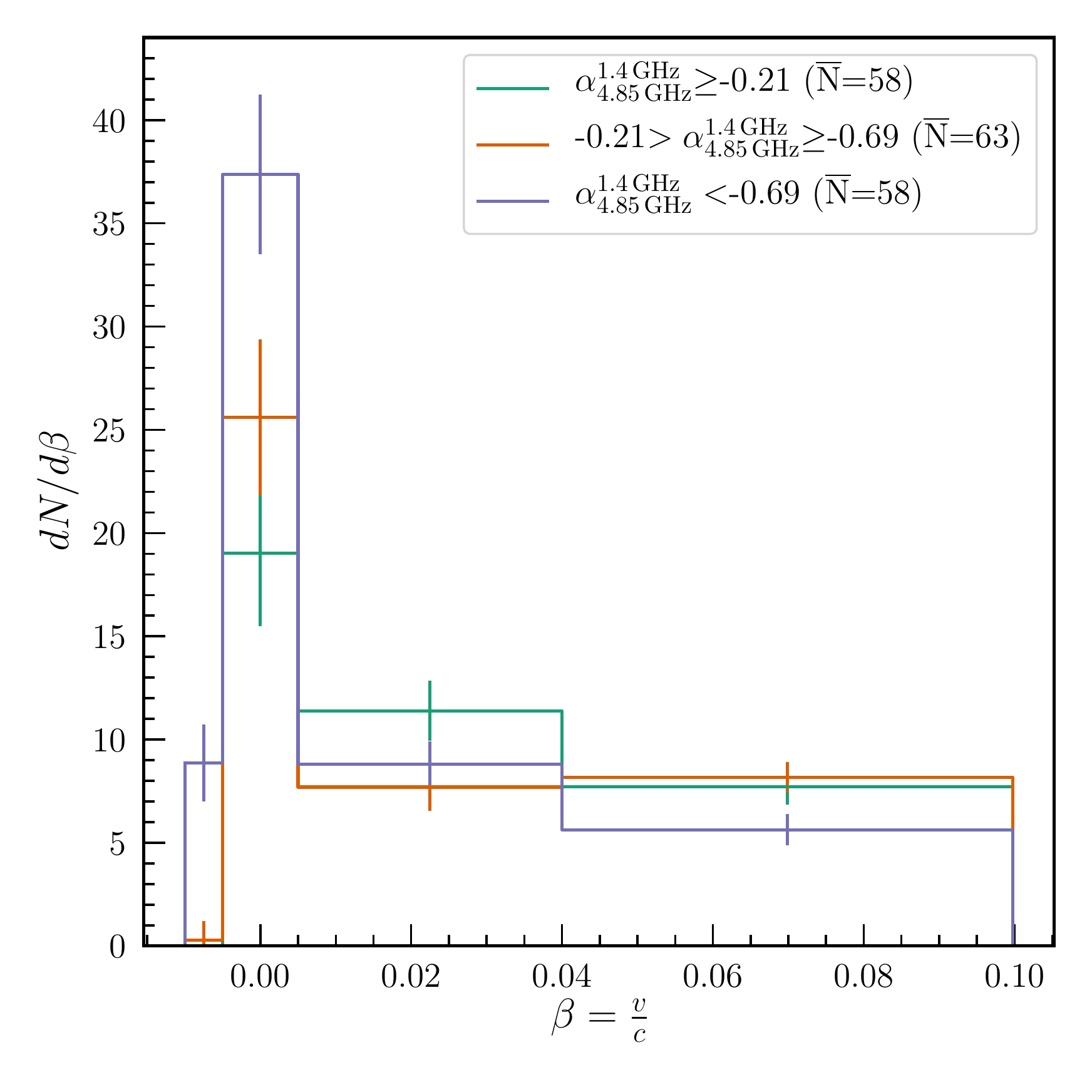}
	\end{subfigure}
	\hfill
	\begin{subfigure}{0.45\textwidth}
		\includegraphics[width=\textwidth]{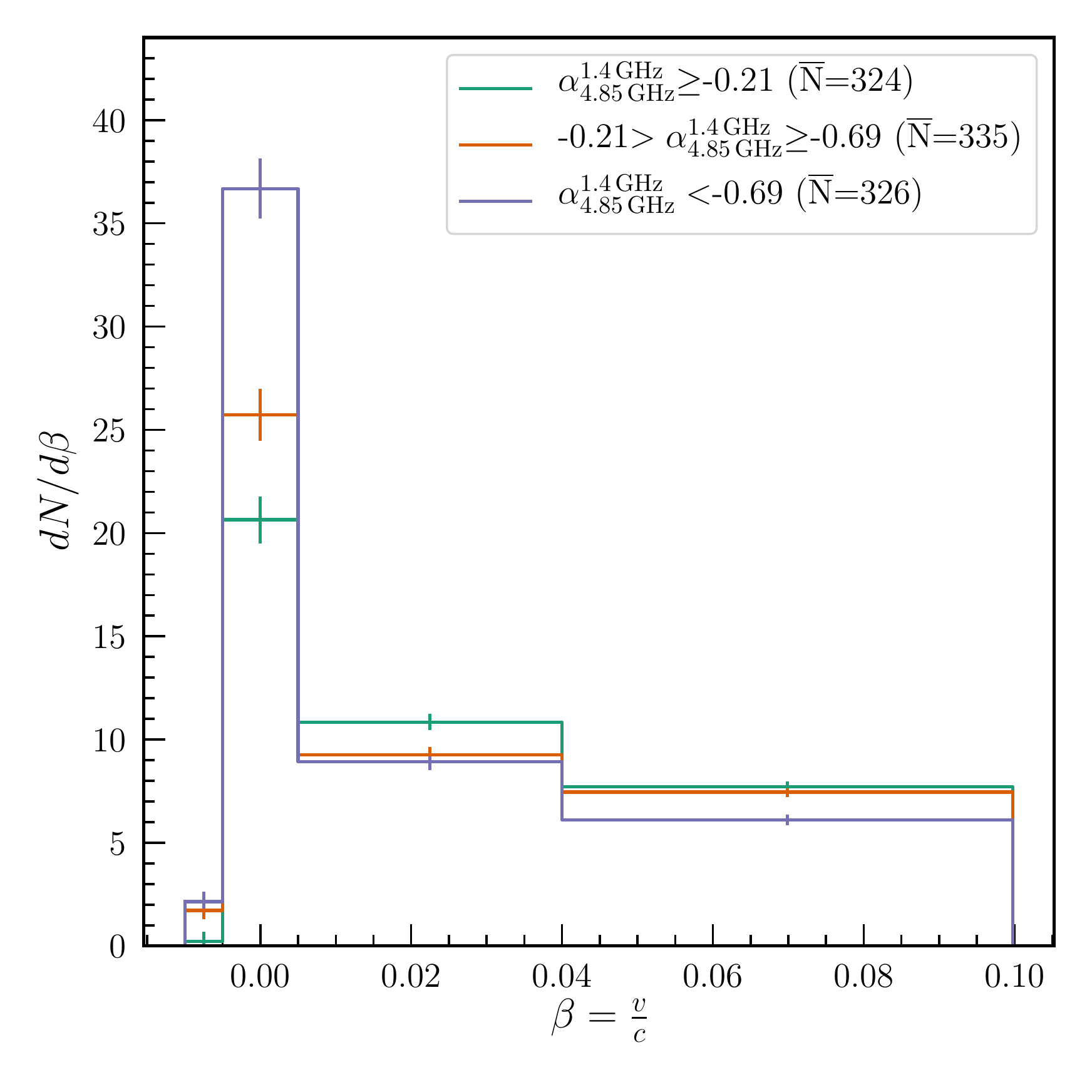}
	\end{subfigure}
    \caption{Average absorber velocity distributions as a function of radio spectral index ($\alpha^{1.4}_{4.85}$) for the matched sub-samples (\emph{left}, $L_{\mathrm{1.4GHz}}$-$z$; \emph{right}, $M_i$-$z$).  We see a consistent trend as observed in Figures~31 \& 32 of \citet{Richards2001b} corroborating our findings in Figure~\ref{fig:alpha20_6_3types}.}
    \label{fig:beta_gtr_alpha_rad}
\end{figure*}

\begin{figure*}
	\centering
	\begin{subfigure}{0.45\textwidth}
		\includegraphics[width=\textwidth]{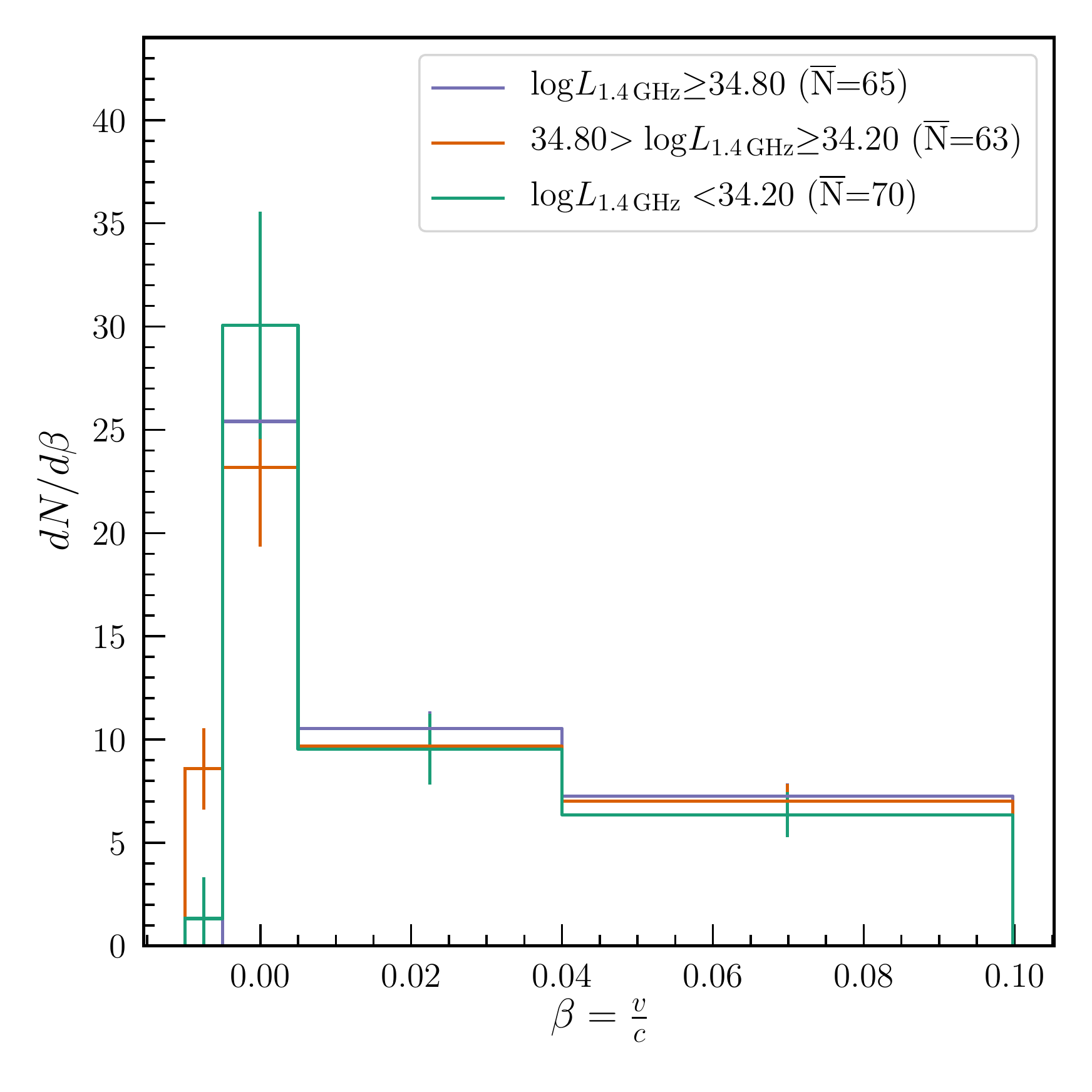}
	\end{subfigure}
	\hfill
	\begin{subfigure}{0.45\textwidth}
		\includegraphics[width=\textwidth]{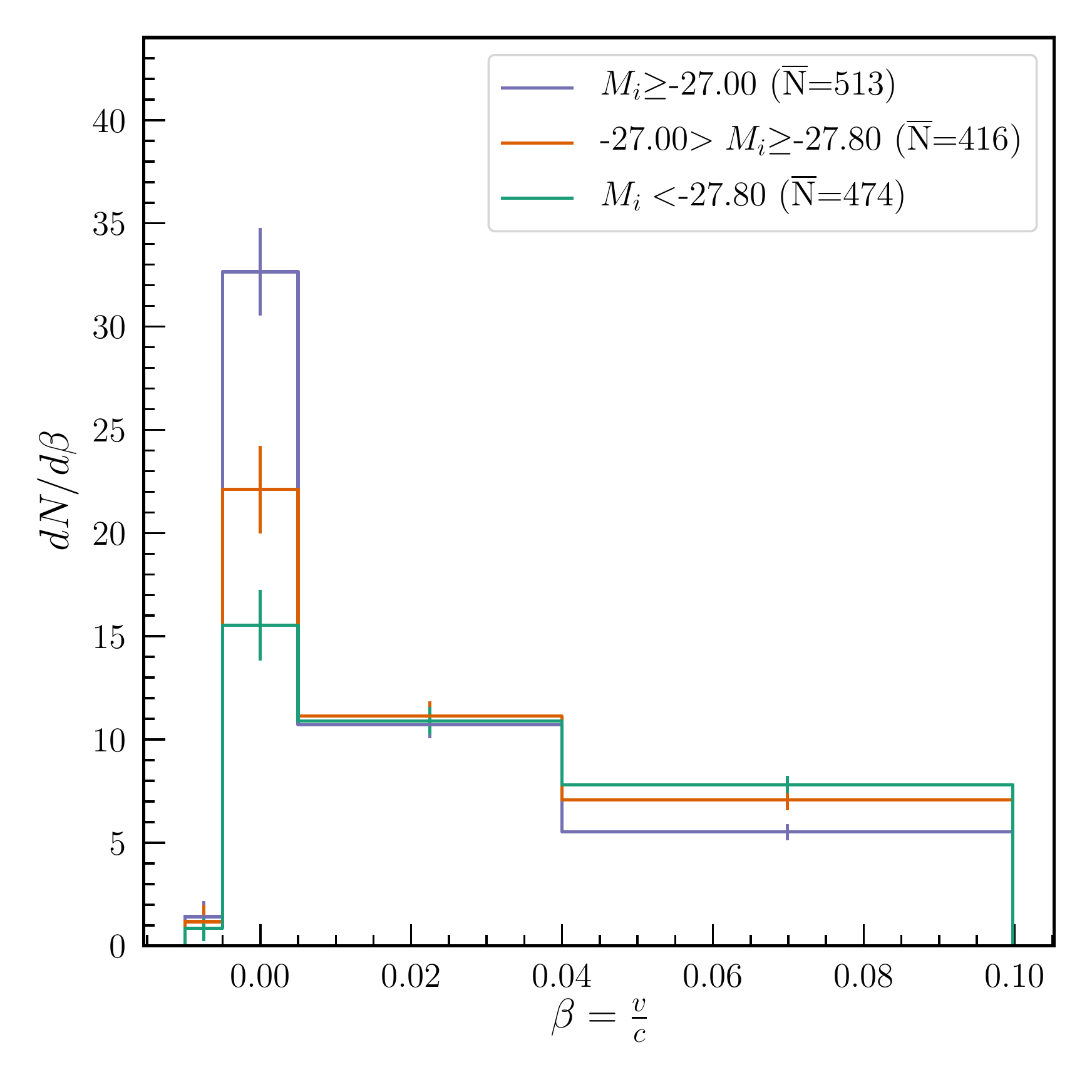}
	\end{subfigure}
	\caption{Average absorber velocity distributions as functions of radio luminosity (\emph{left}) and absolute magnitude (\emph{right}).  Each plot uses the matched sub-sample corresponding to the quasar property by which the absorbers are separated (\emph{left}, $L_{\mathrm{1.4GHz}}$-$z$; \emph{right}, $M_i$-$z$).  Both of these ${\rm dN}/{\rm d}\beta$ plots broadly  agree with their full sample counterparts (Figures~\ref{fig:L_rad} and \ref{fig:abs_mag}, respectively), particularly in that there seems to be no specific absorption trend with radio luminosity, but there is a rather strong correlation of increased associated absorption with decreasing optical luminosity.}  
	\label{fig:beta_gtr_M_L}
\end{figure*}

To obtain the matched samples, we performed a similar process as in Section~\ref{sec:M_i} for making uniform $M_i$ distributions, though now with two parameters instead of just one.  We drew a random sub-sample of the quasars from our main sample in each bin of the 2-D parameter spaces with as many quasars as were in the corresponding bin of the \citet{Richards2001b} sample.  Once we had this matched sub-sample of quasars (1:1 for $L_{\rm rad}$ vs. $z$; 5:1 for $M_i$ vs. $z$), we computed the normalized velocity distributions for the absorbers in the quasars of that sub-sample.  As before we repeated this process 1000 times, taking random sub-samples each time, then took the average of all of the normalized velocity distributions.  Our results for the two 2-D parameter spaces can be seen in Figure~\ref{fig:kde_gtr} (kernel density estimates of the matched sub-samples) and Figures~\ref{fig:beta_gtr_alpha_ro}, \ref{fig:beta_gtr_alpha_rad}, and \ref{fig:beta_gtr_M_L} (average normalized absorber velocity distributions).  The error bars in the plots show a 1$\sigma$ standard deviation from the mean of $dN/d\beta$ for each bin given the 1000 random sub-samples.

We began our comparison by creating ${\rm dN}/{\rm d}\beta$ plots as a function of $\alpha_{\rm ro}$, to compare with the radio loudness results of \citet{Richards2001b}.  We did this for both $L_{\mathrm{1.4GHz}}$-$z$ and $M_i$-$z$ sub-samples and within each, split the quasars into three equal groups based on $\alpha_{\rm ro}$.  In the $L_{\mathrm{1.4GHz}}$-$z$ parameter space, our sample of roughly 2,300 quasars with both radio flux and \ion{C}{IV} absorption measurements was reduced to 110 QSOs, compared to the 143 from \citet{Richards2001b}.  In contrast, the $M_i$-$z$ parameter space, where we have measured apparent magnitudes for nearly all $\sim$ 60,000 quasars in the full absorption sample, showed a closer match between our two samples.  Thus, we were able to include five times as many objects (697) as \citet{Richards2001b} (142; one object did not have a calculated $M_i$) in the matched sample and still maintain an equal $M_i$-$z$ distribution.

Using the $L_{\mathrm{1.4GHz}}$-$z$ matched sub-sample, the left plot of Figure~\ref{fig:beta_gtr_alpha_ro} shows an excess of NALs in louder quasars from low to mid-velocity and no significant trend at high-velocity.  These trends are considerably different than those seen in Figures~28 \& 29 of \citet{Richards2001b}: quieter quasars having a larger excess of high-velocity NALs, and no trend at low-velocity.  Any observed trends all but disappear when the samples are matched in absolute magnitude, suggesting that the radio loudness result in \citet{Richards2001b} (and in Figure~\ref{fig:alpha_ro_wNonRad}) is likely an artifact of sub-samples that are mis-matched in optical/UV luminosity.

Regardless of the method used to match our quasar sample to that in \citet{Richards2001b}, the two plots in Figure~\ref{fig:beta_gtr_alpha_rad} still show an excess of low-velocity intrinsic absorbers in steep-spectrum quasars, thus this result would appear to be robust.  At high velocities, there is an excess of absorbers in flat-spectrum quasars.  Both of these trends are in agreement with Figures~31 \& 32 of \citet{Richards2001b}, which also reveal an excess of low-velocity absorbers in steep-spectrum quasars and of high-velocity absorbers in flat-spectrum quasars. 

In a step beyond direct comparison of the results of \citet{Richards2001b}, Figure~\ref{fig:beta_gtr_M_L} shows the average velocity distributions as functions of $L_{\mathrm{1.4GHz}}$ (\emph{left}) and $M_i$ (\emph{right}) using the corresponding matched sub-samples (Figure~\ref{fig:kde_gtr}, bottom row, \emph{left} and \emph{right} plots respectively).  Even when matched to \citet{Richards2001b} the absorber velocity distribution shows a strong trend in the optical with fainter objects having stronger associated absorption than brighter objects, while the opposite is true at higher velocities.  Furthermore, while the exact shapes of the distributions might be a bit different from our full sample results, the lack of correlation between absorbers as a function of radio luminosity still seems to hold.

To conclude, after restricting our sample of quasars to have similar property distributions as that of \citet{Richards2001b}, we can test some of their findings (and establish the accuracy of our own results). For radio spectral index, we see the same trends as seen by \citet{Richards2001b} as well as in our full sample (Figure~\ref{fig:alpha20_6_3types}), regardless of how the two samples are matched.  This would suggest that the observed radio spectral index trends are real.  On the other hand, the trends in radio loudness seen by \citet{Richards2001b} are not supported by our analysis.  No trend is seen in absorber velocity as a function of radio loudness in our full sample nor in the $M_i$-$z$ matched subsample.

\section{Discussion} \label{sec:discussion}

\subsection{Summary of Observed Results}

Before beginning a discussion on the implications of our results, we first summarize our findings that are most relevant to the discussion.

\begin{table*}
    \footnotesize
    \caption{${\rm d}N/{\rm d}\beta$ as a Function of Velocity Bin}
    \label{tab:summary}
    \begin{tabular}{cccccccccc}
    \hline
    & & \multicolumn{4}{c}{Full Sample} & \multicolumn{4}{c}{Optically Marginalized} \\
    \cline{3-10}
    \multicolumn{2}{c}{Intrinsic Property} & Gaussian & Exponential & Intervening & Figure & Gaussian & Exponential & Intervening & Figure \\
    \hline
	\multirow{2}{*}{FIRST Det.} & Detected & $28.75\pm0.83$ & $9.92\pm0.26$ & $6.39\pm0.16$ & \multirow{2}{*}{Figure~\ref{fig:rad_det}} & \multirow{2}{*}{N/A} & \multirow{2}{*}{N/A} & \multirow{2}{*}{N/A} & \multirow{2}{*}{N/A} \\
	& Undetected & $28.77\pm0.19$ & $10.49\pm0.06$ & $5.96\pm0.04$ & & & & & \\
	\hline
	& ``Lobes" & $32.94\pm1.81$ & $9.46\pm0.52$ & $6.14\pm0.33$ & & & & & \\
	Ext. Rad. & ``Triple" & $29.82\pm1.80$ & $9.27\pm0.54$ & $6.72\pm0.36$ & Figure~\ref{fig:rad_det} & N/A & N/A & N/A & N/A \\
	& ``Core" & $28.45\pm0.93$ & $10.10\pm0.30$ & $6.29\pm0.18$ & & & & & \\
	\hline
	& Louder & $31.20\pm1.58$ & $9.46\pm0.47$ & $6.22\pm0.30$ & & $28.93\pm1.69$ & $9.76\pm0.53$ & $6.36\pm0.33$ & \\
	$\alpha_{\rm ro}$ & Mid & $28.83\pm1.46$ & $9.85\pm0.46$ & $6.51\pm0.29$ & Figure~\ref{fig:alpha_ro_wNonRad} & $27.73\pm1.65$ & $9.99\pm0.53$ & $6.57\pm0.34$ & Figure~\ref{fig:beta_comp_Mi} \\
	& Quieter & $26.68\pm1.30$ & $10.33\pm0.43$ & $6.43\pm0.27$ & & $29.95\pm1.72$ & $10.33\pm0.54$ & $5.92\pm0.32$ & \\
	\hline
	& Louder & $26.14\pm1.40$ & $9.31\pm0.45$ & $7.04\pm0.30$ & & $28.48\pm1.60$ & $9.31\pm0.49$ & $6.64\pm0.32$ & \\
	$L_{\rm 1.4\,GHz}$ & Mid & $26.65\pm1.38$ & $10.44\pm0.46$ & $6.46\pm0.29$ & Figure~\ref{fig:L_rad} & $25.68\pm1.52$ & $10.61\pm0.52$ & $6.49\pm0.32$ & Figure~\ref{fig:beta_comp_Mi} \\
	& Quieter & $33.13\pm1.51$ & $9.99\pm0.44$ & $5.68\pm0.27$ & & $30.54\pm1.66$ & $10.05\pm0.51$ & $5.92\pm0.31$ &\\
	\hline
	& Flat & $22.75\pm2.04$ & $10.49\pm0.75$ & $7.30\pm0.49$ & & $26.56\pm2.80$ & $10.90\pm0.97$ & $6.47\pm0.59$ & \\
	$\alpha_{\rm rad}$ & Mid & $28.65\pm2.30$ & $9.03\pm0.69$ & $6.94\pm0.48$ & Figure~\ref{fig:alpha20_6_3types} & $30.83\pm3.01$ & $8.79\pm0.86$ & $6.79\pm0.60$ & Figure~\ref{fig:beta_comp_Mi}\\
	& Steep & $40.51\pm2.72$ & $8.75\pm0.69$ & $5.43\pm0.43$ & & $38.91\pm3.38$ & $8.74\pm0.86$ & $5.52\pm0.55$ & \\
	\hline
	& Brighter & $20.08\pm0.24$ & $10.89\pm0.09$ & $7.03\pm0.06$ & & & & & \\
	$M_i$ & Mid & $29.27\pm0.31$ & $10.37\pm0.10$ & $5.73\pm0.06$ & Figure~\ref{fig:abs_mag} & N/A & N/A & N/A & N/A \\
	& Fainter & $40.00\pm0.39$ & $9.84\pm0.10$ & $4.75\pm0.06$ & & & & & \\
    \hline
    \end{tabular}
\end{table*}

\begin{itemize}
\item In Figure~\ref{fig:EWr_beta}, we showed that there is an excess of the strongest NALs (in terms of EQW) at low-velocity (i.e., on top of the \ion{C}{IV} emission line), which cannot simply be a S/N feature.  

\item Figure~\ref{fig:zimprov} demonstrates the vast improvement AH19 (and similarly HW10) have made on SDSS quasar systemic redshift measurements, as the gaussian component at $v\sim0$ narrows considerably.  This narrowing reveals that the outflow population is much larger than might otherwise have been expected.

\item The left plot of Figure~\ref{fig:rad_det} shows no difference between absorbers in radio-detected and radio-undetected quasars.

\item In the right plot of Figure~\ref{fig:rad_det}, within the radio-detected group we see moderate changes in the number of low-velocity absorbers as a function of radio morphology.

\item We found no consistent trends in terms of radio loudness as measured by radio-to-optical spectral index and radio luminosity, respectively in Figures~\ref{fig:alpha_ro_wNonRad} and \ref{fig:L_rad}.  Indeed any apparent trends in radio loudness were found to be due instead to underlying trends in optical luminosity (left and middle panels of Figure~\ref{fig:beta_comp_Mi}), which may explain the discrepancy between our results and those of \citet{Richards2001b}.  Accounting for the optical luminosity trend reveals no residual trend in radio loudness.

\item In contrast to the radio loudness results (but consistent with past literature, including \citealt{Richards2001b}),  it was found that radio spectral index correlates strongly with the NAL velocity distribution (see Figure~\ref{fig:alpha20_6_3types}).  The density of ``steep"-spectrum absorbers is nearly double that of ``flat"-spectrum absorbers for low-velocity AALs.   This trend is robust to marginalizing over optical luminosity (see right panel of Figure~\ref{fig:beta_comp_Mi}).

\item The optical luminosity trends are such that less luminous quasars have an excess of low-velocity AALs, but more luminous quasars have an excess of high-velocity NALs (Figure~\ref{fig:abs_mag}). 
  
\item Given that the radio spectral index result is not mitigated by the optical luminosity result, in Figure~\ref{fig:alpha20_6_Mi} we further investigated their combination and showed that low-luminosity, steep-spectrum absorbers have the highest density at low-velocity, while high-luminosity, flat-spectrum absorbers are more prominent at high-velocity.  The differences between these distributions suggest that \SI[separate-uncertainty = true, multi-part-units = repeat]{30.4\pm6.4}{\percent} of high-velocity NALs are intrinsic.
\end{itemize}

\subsection{Implications}

\subsubsection{Evidence for Intrinsic NALs}

Over the years there have been a number of investigations attempting to determine the fraction of NALs (both AALs and at high velocity) that are intrinsic (as opposed to being due to intervening galaxies).  At high velocity, where the intervening hypothesis is the natural default, intrinsic fractions have been estimated at, for example, \SI{36}{\percent} \citep{Richards2001b}, \SIrange[range-phrase=--]{10}{17}{\percent} \citep{Misawa2007}, \SI{45}{\percent} \citep{Wild2008}, \SI{36}{\percent} \citep{Cooksey2013}, and \SI{35.6}{\percent} \citep[][based on the model of \sqcitet{Cooksey2013}]{Bowler2014}.  Based on Figure~\ref{fig:alpha20_6_Mi}, our own analysis suggests that the fraction is \SI[separate-uncertainty = true, multi-part-units = repeat]{30.4\pm6.4}{\percent}.

At lower velocities, including AALs and intermediate velocity, where the intrinsic hypothesis might be assumed, \citet{Wild2008} estimate that 40\% of AALs are intrinsic.  Based on estimates of the areas under the curves, our work shows that improved systemic redshift estimates move a considerable fraction from the gaussian ``possibly associated with the QSO potential well, but not necessarily outflowing" component to the component that is more clearly related to outflows.  Note that AH19 continues the work of HW10 in terms of improving the systemic redshift measurements, and while the AH19 redshifts may remove some residual errors in the HW10 estimates (narrower Gaussian), the number of absorbers in the ``virialized" component remains the same regardless of which redshift estimate is used.  Both, however, demonstrate marked improved over the systemic redshifts taken from the SDSS pipeline).  We conclude that the majority of NALs at $v<\SI{6000}{\km\per\s}$ ($\beta<0.02$) are due to outflows.  Only those NALs left in the narrow ``Gaussian" component could have another origin:  virialized motion around the host quasar rather than outflowing.  Further improvements to the systemic redshifts could continue this trend of a narrower Gaussian component, but this component is unlikely to be eliminated completely.

Potential observational evidence for radiation line-driving (and thus an intrinsic origin) is provided by \citet{Bowler2014} wherein they find signatures of line-locked \ion{C}{IV} absorbers in approximately two-thirds of outflows containing multiple observed absorbers.  Line-locking occurs as the result of inner clouds shielding outer clouds from some of the line-driving radiation that is accelerating them both outwards.  The reduced flux received by the outer clouds from the shielding of the inner clouds traveling at speeds comparable to the separation of the \ion{C}{IV} doublet causes their outflow velocities to synchronize, and produces a triplet in the spectra.  \citet{Bowler2014} found the line-locking phenomenon to occur in BALs and non-BALs roughly equally, which, as they point out, suggests NALs originate from a common source in both BALs and non-BALs.  They show that NALs in outflows extend out to high velocities in both BALs and non-BALs.  In terms of the host quasar SEDs, \citet{Bowler2014} found no evidence of differences in the dust reddening between quasars with outflowing NALs and those with other types of NALs, nor any change in the quasar ``emission" SEDs or luminosity as a function of the presence (or absence) of NALs.

\subsubsection{Accretion Disk Winds and Orientation}

It is one thing to establish that some fraction of NALs are intrinsic; it is another thing to establish a physical model for such systems, and in the introduction we detailed various different forms of AGN feedback.  A number of authors have argued for an accretion disk wind model for some fraction of intrinsic NALs (potentially as ``failed" BALs), with orientation playing a role in the probability of their appearance \citep[e.g.,][]{Hamann1997,Vestergaard2003,Misawa2007,Wild2008}.  Determining if the empirical results presented herein point to an answer about which form(s) of feedback are in operation in NAL quasars is beyond the scope of this paper. Nevertheless, we highlight  some aspects of our results that may be consistent with radiation line-driven winds coupled with orientation effects, while recognizing that other explanations are certainly possible (e.g., kiloparsec-scale winds associated with star formation \sqcitep{Barthel2017}).

First, Figure~\ref{fig:EWr_beta} shows that it is clear that there is an excess of strong AALs (at $v\sim0$) since any bias in terms of finding NALs should go in the other way (smaller EQW more likely on top of high S/N emission lines).  This figure suggests that the AAL population is real, and particularly strong.  Second, when splitting the absorbers by optical/UV luminosity (or absolute magnitude), as in Figure~\ref{fig:abs_mag}, we found two clear trends: (1) the least luminous quasars in our sample have a significant excess of AALs as compared to the most luminous quasars, and (2) the most luminous quasars in our sample show an excess of high-velocity NALs relative to lower luminosity quasars.  These observations possibly suggest that higher luminosity quasars are better able to accelerate the gas clouds that produce absorption features, such as in the model of \citet{Murray1995} where higher luminosity is associated with a relatively lower X-ray luminosity, enabling more efficient line driving.

If we accept the idea that quasar orientation causes changes in radio spectral indices, and if NALs are dominated by rotational and polar motion (rather than purely equatorial) then we might expect to see an excess of low (or 0) velocity systems in steep-spectrum quasars which are more ``edge on", as the motion of the absorption line material is seen in projection.  If that material is seen at a larger angle from the disk (as indicated by a flatter radio spectrum), then more of the absorber's outflow velocity would be seen along our line of sight.  These results can be seen in Figure~\ref{fig:alpha20_6_3types}. We note that for the radio detections at the redshifts investigated herein the radio luminosity is generally much higher than the luminosity where FRIIs \citep{Fanaroff1974} dominate, so it seems unlikely that radio spectral index differences between FRI and FRII radio sources play a role (unless the FIRST-undetected quasars turn out to be FRI).

Both Figure~\ref{fig:alpha20_6_Mi} and the far right panel of Figure~\ref{fig:beta_comp_Mi} suggest that the strong effect seen in the radio spectral index is an effect independent from the luminosity effect, possibly consistent with an accretion disk wind model modified by orientation.  Appropriate coupling of the radio spectral index with optical/UV luminosity produces an even more extreme effect on the AAL population (as well as to the high velocity population as discussed above).  Similar effects were seen in \citet{Richards2001b}.  The excess of AALs and relative lack of high-velocity NALs in steep-spectrum, faint quasars could be consistent with an object that is seen through a ``failed" wind (or that simply is less suited to driving a strong wind) and that is seen edge-on, creating a ``pile up" at low velocity.  On the other hand, the lack of AALs, but excess of high-velocity NALs in flat-spectrum, luminous quasars may be consistent with an object that is capable of driving a strong wind from the accretion disk and is seen more along the outflow velocity vector.

Alternatively, it may be more likely that these trends are not due to projection effects in a universal population of absorbers, but rather indicate lines of sight through different parts of an outflow \citep[e.g., Figure 1 of][]{Proga2003}.  Specifically, for high-$L$, flat-spectrum objects, the absorbers might be viewed through the higher velocity part of the wind that is closer to disk normal.  Conversely, in low-$L$, steep-spectrum objects the absorbers might be viewed through lines of sight at larger angles from disk normal that pass through a lower velocity component of the wind.  This interpretation may be consistent with Figures~7 and 8 of \citet{Bowler2014}, where some high-ionization lines, like \ion{N}{V}, become weaker with increasing velocity (suggesting larger distances), while some low-ionization lines, like \ion{Mg}{II}, become weaker with decreasing velocity (suggesting absorbers closer to the central engine).  We note that the SDSS-DR7 absorbers investigated herein are the same ones in the \citet{Bowler2014} study and show evidence of line-locking in both low- and high-velocity NALs, possibly providing further support for a radiation line driving interpretation.

Beyond the hints from ionization states as a function of velocity seen in \citet{Bowler2014}, our work herein does not provide any direct insight into the question of the radial location of the absorbers.  However, we acknowledge that knowing the distance of the absorbers (both broad and narrow) from the central engine is crucial to understanding the nature of intrinsic NALs \citep[e.g.,][]{Barthel2017, Dabbieri2018, Arav2018}.

One avenue for future work is that if strong AALs indicate edge-on orientation, their presence could be used as an orientation indicator for {\em radio-quiet} quasars.  In that regard, it would be interesting to look for differences in the emission lines of both RL and RQ quasars with strong AALs.  If the emission line properties are correlated with the (in)ability to drive a strong wind \citep[e.g.,][]{Richards2011}, then we might expect that both RQ and RL quasars with strong AALs will have similar emission line properties--such as strong \ion{C}{IV} emission and redder continua.  If that is true, then we might expect that RQ quasars lacking strong AALs may exhibit statistically different emission line properties as compared to RQ quasars with strong AALs.  Alternatively, the combination of AALs and steep radio spectral index could both be related to ``frustrated" outflows as might be the case in compact steep-spectrum sources rather than primarily an indication of orientation \citep{ODea1998}.

\subsubsection{Origin of Radio Emission} \label{sec:disc_rad}

Part of the impetus for this work involved asking how the population of narrow absorption lines in quasars are empirically related to the radio properties of the quasars that host those NALs.  The simplest test was to see if the detection (or lack thereof) of radio flux affected the NAL distribution.  Although the excess of AALs in steep-spectrum RL quasars is well-known \citep[e.g.,][]{Foltz1986}, this fact has potentially given the impression that AALs are a phenomenon of radio-loud quasars.   Our results suggest that neither radio detection probability, nor the two standard measures of radio loudness affect the NAL population---a result that both has support and contradictions in the literature.

Assuming that most previous differences seen in the NAL population between RL and RQ quasars were due to selection effects (such as optical/UV luminosity) and that there is no inherent difference between the RL and RQ populations, suggests an interesting interpretation.   Namely, that the radio properties of quasars are effectively stochastic.  Specifically, the randomness of radio with respect to absorption could be consistent with a magnetically-arrested disk (MAD) model \citep{Narayan2003, Sikora2013, Rusinek2017}.  In such a model, the radio may be more related to the build-up of magnetic flux in the gap region than with anything happening further out in the accretion disk.  Possibly the magnetic flux could even have been accumulated during a prior merger.  Thus, the presence of radio emission could have little to do with the current accretion cycle and might explain why the presence of strong radio emission appears largely random.  

Specifically, \citet[][hereafter SB13]{Sikora2013} suggest that magnetic flux may better explain the large range in AGN jet power than spin and/or Eddington ratio.  This requires accumulation of significant magnetic flux around the black hole which is difficult in standard thin disk models, but which SB13 argue might be solved by a 2nd generation cold merger (yielding the current incarnation of the AGN) following a 1st generation hot merger, trapping the magnetic flux in the gap region between the disk and the BH.

Differences in the population in terms of likelihood of a second generation merger could lead to radio not being completely random.  For example, the concept of an accretion-rate magnetic flux dependence to the formation of radio jets in AGNs \citep{Rusinek2017} might help explain why the RL fraction is a function of luminosity \citep{Jiang2007} with luminous quasars being radio loud at $<5$\% \citep{Kratzer2015}, but  lower-luminosity AGNs at much higher rates \citep[e.g.,][]{Best2005}.  Simply put, luminous quasars would need a larger amount of magnetic flux in the gap region to produce jets.

In this context, \citet{Zamfir2008} find ambiguous results in regards to the existence of a RQ--RL dichotomy, particularly when the radio is compared to bolometric luminosity.  Thus, it is interesting to consider the claim of \citet{Kratzer2015} that one should not make two-way RL vs.\ RQ comparisons, but rather split the RQ population into ``hard-spectrum RQ" (HSRQ), which are more like RL (presumably with lower $L/L_{\rm Edd}$) and ``soft-spectrum RQ" (SSRQ), which likely have higher $L/L_{\rm Edd}$.  In the accretion disk wind model, HSRQs might be expected to have more AALs, while SSRQs might be expected to have more high-velocity NALs---due to different abilities to drive a wind (independent of orientation).

We suggest that the similarity of the AAL population between RL and RQ and the similarity of HSRQ and RL could both be related to the MAD process as envisioned by SB13.  Specifically, the existence of FIRST-detected radio emission in $\sim 4\%$ of quasars \citep{Kratzer2015} could be due to the existence of a previous generation hot merger in these objects.  The likelihood of such a merger may be essentially stochastic among the quasar population, but might be somewhat higher for sources in denser environments (explaining the discrepancy in the radio-loud fraction between HSRQ and SSRQ).  If AAL systems are unrelated to the radio emission, but instead are related to the efficiency of driving an accretion disk wind, one might expect the AAL distribution of RL and HSRQ to be similar, with the presence of strong AAL systems in HSRQ being indicative of edge-on orientation in the same way that steep-spectrum radio spectral index indicates edge-on orientation in RL sources.

Although this paper highlights our results in the light of a radiation-driven accretion disk wind, we refer the reader back to Section~\ref{sec:intro} for other potential outflow mechanisms. Furthermore, we note that in our interpretation the similarity of the NAL distributions in RL and RQ quasars suggests that the NALs are unrelated to the jet, or at least that, as suggested by \citet{Tombesi2014}, ``the presence of relativistic jets does not preclude the existence of winds".  However, in the case where the jets are confined \citep[e.g.,][]{Begelman1989}, frustrated \citep{ODea1998}, and/or are simply undetected at the limit of the FIRST survey, it could still be that there is a significant relationship between jets and NALs in both types of quasars.

\section{Conclusions} \label{sec:conclusions}

One of the largest data samples of \ion{C}{IV} narrow absorption line systems was analyzed to determine relationships between intrinsic quasar properties and intrinsic NALs.  With the help of improved systemic quasar redshift measurements from AH19, it was determined that the fraction of high velocity NALs ($\gtrsim \SI{12000}{\km\per\s}$) that are intrinsic is roughly \SI[separate-uncertainty = true, multi-part-units = repeat]{30.4\pm6.4}{\percent}---comparable to previously reported estimates in the literature.

As others have noted in prior literature, this work finds an excess of AALs in steep-spectrum quasars compared to flat-spectrum quasars.  These findings may be consistent in the context of the radio spectral index being an indicator of quasar line-of-sight orientation.  Steep-spectrum quasars (seen more edge-on in an orientation model) observe the wind either in projection or simply through a lower-velocity part of the outflow (that is located at angles farther from disk normal), while flat-spectrum quasars (seen more face-on) observe the outflow along its line-of-sight or through a higher-velocty part of the outflow (closer to disk normal).  Although the goal of this work is not to prove or disprove any particular physical model for NALs, when the radio spectral index and optical/UV luminosity results are combined, trends are produced that are potentially consistent with an accretion disk wind model, modified by orientation.  A significant excess of AALs in faint, steep-spectrum quasars could indicate objects with weak or ``failed" winds viewed nearly edge-on.  On the other hand, the excess of high velocity NALs in luminous, flat-spectrum quasars could be the result of a quasar capable of driving a wind to high speeds and viewing the NAL outflow nearly along its direction of motion.

Finally, the lack of correlation between radio properties and NAL outflows is discussed (particularly in how this contradicts previous results in the literature).  After marginalizing the radio properties over absolute magnitude, any previously observed trends all but disappear, suggesting that the differences between RL and RQ NAL populations are due to selection effects and are not physical.  This leads to an interesting interpretation:  quasar radio properties are stochastic in nature.  Specifically, the random radio results with respect to absorption could be consistent with a magnetically-arrested disk model (that is largely unconnected with the physics of NAL/BAL outflows).

\section*{Acknowledgements}
We would like to thank Paul Hewett for his insight, comments, and discussions, as well as for the use of his \ion{C}{IV} NAL catalogs and preliminary SDSS DR12 quasar redshift estimates.  We would also like to thank Nadia Zakamska for her discussion and thoughtful insights, especially regarding wind model interpretations.  Finally, we'd like to thank the referee for their helpful comments and suggestions.

Funding for the SDSS and SDSS-II has been provided by the Alfred P. Sloan Foundation, the Participating Institutions, the National Science Foundation, the U.S. Department of Energy, the National Aeronautics and Space Administration, the Japanese Monbukagakusho, the Max Planck Society, and the Higher Education Funding Council for England. The SDSS Web Site is http://www.sdss.org/.

The SDSS is managed by the Astrophysical Research Consortium for the Participating Institutions. The Participating Institutions are the American Museum of Natural History, Astrophysical Institute Potsdam, University of Basel, University of Cambridge, Case Western Reserve University, University of Chicago, Drexel University, Fermilab, the Institute for Advanced Study, the Japan Participation Group, Johns Hopkins University, the Joint Institute for Nuclear Astrophysics, the Kavli Institute for Particle Astrophysics and Cosmology, the Korean Scientist Group, the Chinese Academy of Sciences (LAMOST), Los Alamos National Laboratory, the Max-Planck-Institute for Astronomy (MPIA), the Max-Planck-Institute for Astrophysics (MPA), New Mexico State University, Ohio State University, University of Pittsburgh, University of Portsmouth, Princeton University, the United States Naval Observatory, and the University of Washington.

Funding for SDSS-III has been provided by the Alfred P. Sloan Foundation, the Participating Institutions, the National Science Foundation, and the U.S. Department of Energy Office of Science. The SDSS-III web site is \url{http://www.sdss3.org/}.

SDSS-III is managed by the Astrophysical Research Consortium for the Participating Institutions of the SDSS-III Collaboration including the University of Arizona, the Brazilian Participation Group, Brookhaven National Laboratory, Carnegie Mellon University, University of Florida, the French Participation Group, the German Participation Group, Harvard University, the Instituto de Astrofisica de Canarias, the Michigan State/Notre Dame/JINA Participation Group, Johns Hopkins University, Lawrence Berkeley National Laboratory, Max Planck Institute for Astrophysics, Max Planck Institute for Extraterrestrial Physics, New Mexico State University, New York University, Ohio State University, Pennsylvania State University, University of Portsmouth, Princeton University, the Spanish Participation Group, University of Tokyo, University of Utah, Vanderbilt University, University of Virginia, University of Washington, and Yale University.

%%%%%%%%%%%%%%%%%%%%%%%%%%%%%%%%%%%%%%%%%%%%%%%%%%

%%%%%%%%%%%%%%%%%%%% REFERENCES %%%%%%%%%%%%%%%%%%

\bibliographystyle{mnras}
\bibliography{abs_rad}

\bsp	% typesetting comment
\label{lastpage}
\end{document}